\documentclass[pmlr,twocolumn]{jmlr}
\hypersetup{urlcolor={blue}}
\usepackage{longtable}
\usepackage{booktabs}
\usepackage[load-configurations=version-1]{siunitx}
\jmlrvolume{193}
\jmlryear{2022}
\jmlrworkshop{Machine Learning for Health (ML4H) 2022}

\title[A Path Towards Clinical Adaptation of Accelerated MRI]{A Path Towards Clinical Adaptation of Accelerated MRI}

\author{
  \Name{Michael S. Yao\nametag{\thanks{Work done during an internship at\newline Microsoft Research.}}}\Email{michael.yao@pennmedicine.upenn.edu}\\
  \addr Microsoft Research \\ University of Pennsylvania, Department of Bioengineering \\ University of Pennsylvania, School of Medicine \\
  \AND
  \Name{Michael S. Hansen}\Email{michael.hansen@microsoft.com}\\
  \addr Microsoft Research
}

\begin{document}
\maketitle

\begin{abstract}
Accelerated MRI reconstructs images of clinical anatomies from sparsely sampled signal data to reduce patient scan times. While recent works have leveraged deep learning to accomplish this task, such approaches have often only been explored in simulated environments where there is no signal corruption or resource limitations. In this work, we explore augmentations to neural network MRI image reconstructors to enhance their clinical relevancy. Namely, we propose a ConvNet model for detecting sources of image artifacts that achieves a classifier \textit{F\textsubscript{2}} score of 79.1\%. We also demonstrate that training reconstructors on MR signal data with variable acceleration factors can improve their average performance during a clinical patient scan by up to 2\%. We offer a loss function to overcome catastrophic forgetting when models learn to reconstruct MR images of multiple anatomies and orientations. Finally, we propose a method for using simulated phantom data to pre-train reconstructors in situations with limited clinically acquired datasets and compute capabilities. Our results provide a potential path forward for clinical adaptation of accelerated MRI.
\end{abstract}

\begin{figure*}[ht]
  \floatconts
    {fig:paper-overview}
    {\caption{Overview of a standard MR pipeline, where $k$-space signal data is acquired from a scanner environment, pre-processed, and used to reconstruct an MR image. At each step, we highlight limiting issues in real-world applications that our contributions individually seek to address.}}
    {\includegraphics[width=\textwidth]{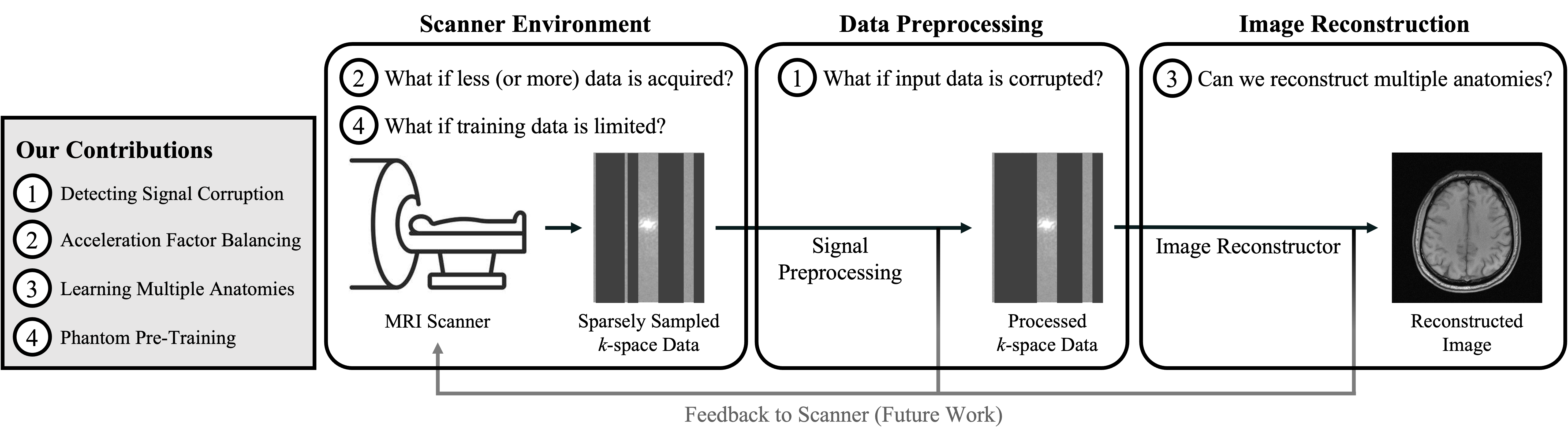}}
  \end{figure*}

\section{Introduction}
\label{section:introduction}

Magnetic resonance imaging (MRI) is a powerful noninvasive imaging modality used to diagnose a wide variety of patient diseases. Compared to other structural imaging modalities, MRI does not require patient irradiation, unlike computed tomography and X-ray techniques, and has superior resolution compared to ultrasound imaging. However, MRI is rarely used as a first-line diagnostic tool due to its high cost and long scan times. These limitations can often prevent certain patient groups from taking advantage of the clinical benefits of MRI.

To address these issues, various techniques have been developed to reconstruct MR images from sparsely acquired data and broadly fall under the domain of \textbf{accelerated MRI}. These methods include compressed sensing, simultaneous multislice imaging, and more recently, neural networks \citep{cs-review, tmv-cs, multislice}. Of note, the introduction of fastMRI, MRNet, and other large MR datasets have enabled recent work on training deep learning models to reconstruct images from undersampled data \citep{zbontar2018fastMRI, mrnet}.

While much success has been found in developing novel algorithms and approaches in accelerated MRI reconstruction, the vast majority of these methods are developed in isolated sandbox environments that do not take hardware and clinical constraints into account. For example, current reconstructor neural networks are trained on often prohibitively large datasets for single-anatomy reconstruction tasks with little environmental signal contamination or patient motion \citep{bakker, sriram, loupe}. Such datasets also often have a fixed fraction(s) of signal data acquired, which is not representative of the range of undersampling seen by a general-purpose MRI reconstructor.

In this work, we attempt to overcome these limitations by adjusting model datasets, loss functions, and training pipelines. In doing so, we hope to provide a path forward towards clinical adaptation of learning-based methods in accelerated MRI reconstruction. Our contributions are as follows:
\begin{enumerate}
    \item (\textbf{Detecting Signal Corruption}) We introduce a self-supervised $k$-space data corruption detector specifically formulated for sparsely sampled MR datasets and show that it accurately flags signal data adversely affected by simulated patient motion and environmental noise.
    \item (\textbf{Acceleration Factor Balancing}) We explore the benefits of training reconstruction models on variably accelerated data to more accurately simulate the range of undersampling seen by MR scanners.
    \item (\textbf{Learning Multiple Anatomies}) We propose a loss function to overcome catastrophic forgetting in sequentially learning to reconstruct knee and brain datasets, and investigate tradeoffs in model performance within these two domains.
    \item (\textbf{Phantom Pre-Training}) We offer a framework for using transfer learning with simulated phantom data to train reconstruction models even when there is limited clinical data available.
\end{enumerate}
We break down remaining sections on related work (\sectionref{section:related-work}), methods (\sectionref{section:methods}), results (\sectionref{section:results}), and discussion (\sectionref{section:discussion}) by the above contributions.\footnote{In navigating our work, we recommend reading by contribution as opposed to linearly by numbered section. For example, reading the Background and Related Work, Methods, and Results associated with `Detecting Signal Corruption' together before moving onto subsections related to another contribution.}

\section{Background and Related Work}
\label{section:related-work}
\subsection{Detecting Signal Corruption}
\label{section:discriminator-related-work}
One of the main limitations to effective image reconstruction is the presence of noise and signal contamination from sources such as the patient, the internal scanner electronics, and the local clinical environment. To mitigate these adverse contributions to the final reconstruction, \citet{shaw2020} developed a computational model for motion artifacts and used it to remove artifacts from MR images. However, using such a method would require intermediate image reconstructions that are inherently tied to the choice of accelerated MRI reconstructor. Works by \citet{bydder2002, loktyushin2012, johnson2018}, and \citet{usman2020} correct for motion working in $k$-space alone, but have only been validated for fully-sampled $k$-space datasets and can therefore solely be used retrospectively after the signal acquisition process has already finished. This limits their potential use cases in accelerated MRI.

Our ideal preprocessing module is capable of online, real-time signal denoising after each acquisition step. Furthermore, its performance should be independent of both the scanner environment and the proportion of available raw data at any stage of the MR scan. Given these constraints, we trained a signal corruption detector to compare newly acquired $k$-space lines with previously acquired Fourier data in order to predict whether the new lines will need to be re-acquired in subsequent acquisition steps.

\subsection{Acceleration Factor Balancing}
\citet{hammernik} introduced the Variational Network (VarNet) reconstructor model, which can be thought of as an iterative unrolling of compressed sensing algorithms that attempt to reconstruct images subject to a regularization constraint, such as total variation minimization. \citet{sriram}, \citet{bakker}, and others built upon this work by co-learning sensitivity map estimation (E2E-VarNet) to achieve state-of-the-art performance on the fastMRI challenge leaderboard and when compared to alternative reported architectures, such as U-Net's \citep{yin2021end, loupe, zbontar2018fastMRI, unet}. Each of these works defined the \textbf{acceleration factor} for each training dataset as the inverse of the proportion of acquired data.

In the past, accelerated MRI models were trained and assessed on undersampled $k$-space datasets representing a finite space of acceleration factors, such as 4x and 8x. However, signal preprocessing steps, such as the one described in \sectionref{section:discriminator-related-work}, may be used to throw out corrupted $k$-space lines that were acquired, resulting in acceleration factors and pseudo-random sampling masks that these models may not have seen previously. Furthermore, it may be difficult to determine how long a patient is able to tolerate the MR scanning environment ahead of time, and so more or less data may be available at reconstruction time. Given these reasons and the sensitivity of deep learning methods in accelerated MRI reconstruction to perturbations in acceleration factors during inference reported by \citet{mrinstability}, we sought to train an E2E-VarNet on undersampled $k$-space data with variable acceleration factors and random sampling patterns to better capture the space of possible MR sampling masks that could be fed into an accelerated MRI reconstructor.

\subsection{Learning Multiple Anatomies}
Although deep learning-based reconstruction methods have demonstrated promising results, the number of parameters and consequent model size often prove unwieldy in production pipelines. Especially given that the standard MR scanner must be able to image any relevant clinical anatomy, training and storing different models for different anatomies, orientations, and or scanning sequences can prove intractable in downstream applications to patient care. It would be advantageous to train a \textit{single} model to reconstruct images from multiple datasets, such as those of both knee and brain anatomies. Such a `general-purpose' reconstructor model may also be useful in pre-training reconstructors that could then be fine-tuned on smaller, population-specific datasets for downstream clinical use, as previously explored in tasks such as medical image classification and segmentation \citep{radimagenet}.

Sequential learning of multiple tasks has been previously explored in other domains outside of accelerated MRI reconstruction \citep{glioma-ewc}. One of the main limitations in these works is \textit{catastrophic forgetting} where a model will forget how to perform a previously learned task while training on a separate task. \citet{catastrophic-forgetting} proposed a method to overcome this phenomenon, termed \textit{elastic weight consolidation} (EWC), that involves a regularization penalty term contribution to the loss function during model training. While the benefits of EWC have been explored across different domains, the applications of EWC remain to be investigated in accelerated MRI reconstruction. Here, we define a loss function to train a single reconstructor model to sequentially learn to reconstruct undersampled knee, and then brain, datasets.

\subsection{Phantom Pre-Training}
Deep learning methods are a powerful tool for reconstructing MRI images from sparsely sampled data. However, traditional approaches using neural networks require large amounts of data to adequately train a reconstructor model. Given stringent patient confidentiality requirements and the high cost of clinical MRI scans, acquiring sufficient data is a major bottleneck in signal and image processing research. Large datasets are also often unwieldy to manage and expensive to store, further compounding the compute costs and resource limitations that retard progress in engineering new methods and applying existing ones to clinical use cases.

Recent advancements in transfer learning have made it possible to transfer knowledge learned from one task and apply it to another \citep{tl-review, truong2021, tl-nips}. While prior work has primarily focused on applications of transfer learning for medical image classification tasks, we hypothesized that pre-training a reconstructor model on simulated Shepp-Logan brain phantom data described by \citet{shepplogan} and then fine-tuning model weights on clinically acquired data could still yield promising model performance.

\section{Methods}
\label{section:methods}
\paragraph{Dataset} We trained our models on the fastMRI dataset, a publicly available set of knee and brain MR scans from varying anatomical orientations and MR pulse sequences, including proton-weighted, T1-weighted, and T2-weighted. The dataset offers $41,877$ ($92,950$) fully sampled multi-coil knee (brain) slices from $1,172$ ($5,847$) volumes, of which $34,742$ ($70,748$) slices from $973$ ($4,469$) volumes were used for training. Validation and test set partitions varied by experiment and are detailed in \sectionref{section:implementation-details}.

\paragraph{Model Training} We trained our models using an Adam optimizer with $\beta$-parameters of $\beta_1=0.9, \beta_2=0.999$ and a learning rate of $\eta=10^{-3}$ that decreased ten-fold every $40$ epochs. Models were trained for a total of $50$ epochs unless otherwise noted using 8 NVIDIA Tesla V100 GPUs.

\subsection{Detecting Signal Corruption}
To detect lines of $k$-space that may have been corrupted by common clinical sources, such as simulated patient motion or $k$-space spiking, we implemented a convolutional neural network (ConvNet) $f$ to identify corrupted $k$-space lines from sparsely sampled MR data (\figureref{figure:cnn-architecture-text}). $f:\mathbb{R}^{640\times 2}\rightarrow \mathbb{R}$ is an approximately $20$ K-parameter network with two convolutional/max-pooling layers followed by two fully connected layers, all with Leaky ReLU activation. The input to the ConvNet is the magnitude of a high-frequency line $\mathbf{k}_h$ of $k$-space that we seek to determine whether it is corrupted, and the magnitude of a reference low-frequency line $\mathbf{k}_{\ell}$ assumed to be nominally uncorrupted. $\mathbf{k}_{\ell}$ is a member of the \textit{auto-calibration signal} (ACS), which is defined as the center fraction of the acquired signal data and assumed to be nominally uncorrupted. Conveniently, the ACS can be programmed to be the first set of lines to be acquired in an MR scan for tasks such as sensitivity map estimation \citep{bakker, sriram}.

The final output of our model $\tilde{y}_h\in\mathbb{R}$ between $0$ and $1$ describes the model's predicted likelihood that the input $k$-space line was corrupted. Our model was trained using the binary cross entropy loss function.

\begin{figure}[h]
  \floatconts
    {figure:cnn-architecture-text}
    {\caption{Signal corruption detector model $f$. The model outputs were averaged over the auto-calibration signal (ACS) to yield the final model output for an acquired line $\mathbf{k}_h$.}}
    {\includegraphics[width=0.48\textwidth]{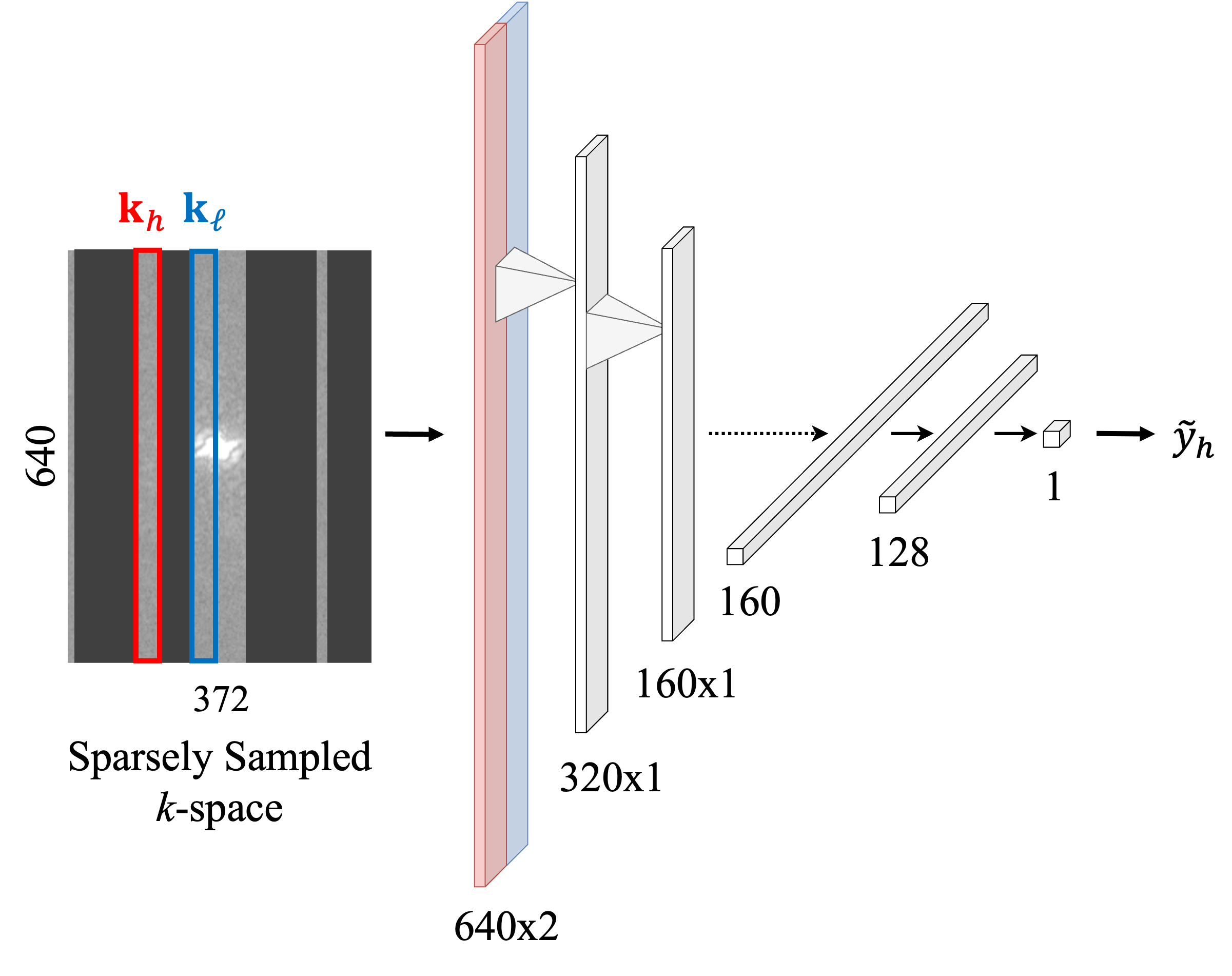}}
\end{figure}

\subsection{Acceleration Factor Balancing}
\label{section:var-acceleration-methods}
To explore the impact of training reconstructor models on variably accelerated data, we used the E2E-VarNet reconstructor model consisting of $8$ cascade layers constituting $20.1$ M trainable parameters. \citet{sriram} describes the details of the E2E-VarNet implementation. Models were trained to maximize the structural similarity index measure $\textnormal{SSIM}(\tilde{\mathbf{x}}, \mathbf{x})$ between a predicted image $\tilde{\mathbf{x}}$ and the associated ground truth $\mathbf{x}$ from the fully sampled dataset. Images were cropped to a final size of $256\times 256$ in computing the model loss following the convention of \citet{yin2021end, bydder2002}, and others. For quantitative evaluation of our reconstructor models, we evaluated the SSIM, normalized mean square error (NMSE), and peak signal-to-noise ratio (PSNR).

\begin{figure*}[ht]
  \floatconts
    {fig:discriminator-graphs}
    {\caption{Discriminator ConvNet (a) heatmap distributions, (b) receiver operating characteristic (ROC) curve, and (c) classifier metrics on hold-out test dataset. Discussion on qualitative results is provided in \sectionref{section:discriminator-qualitative} and \figureref{fig:discriminator-images}.}}
    {
      \subfigure[][b]{\label{fig:detector-heatmap}\includegraphics[width=0.33\textwidth]{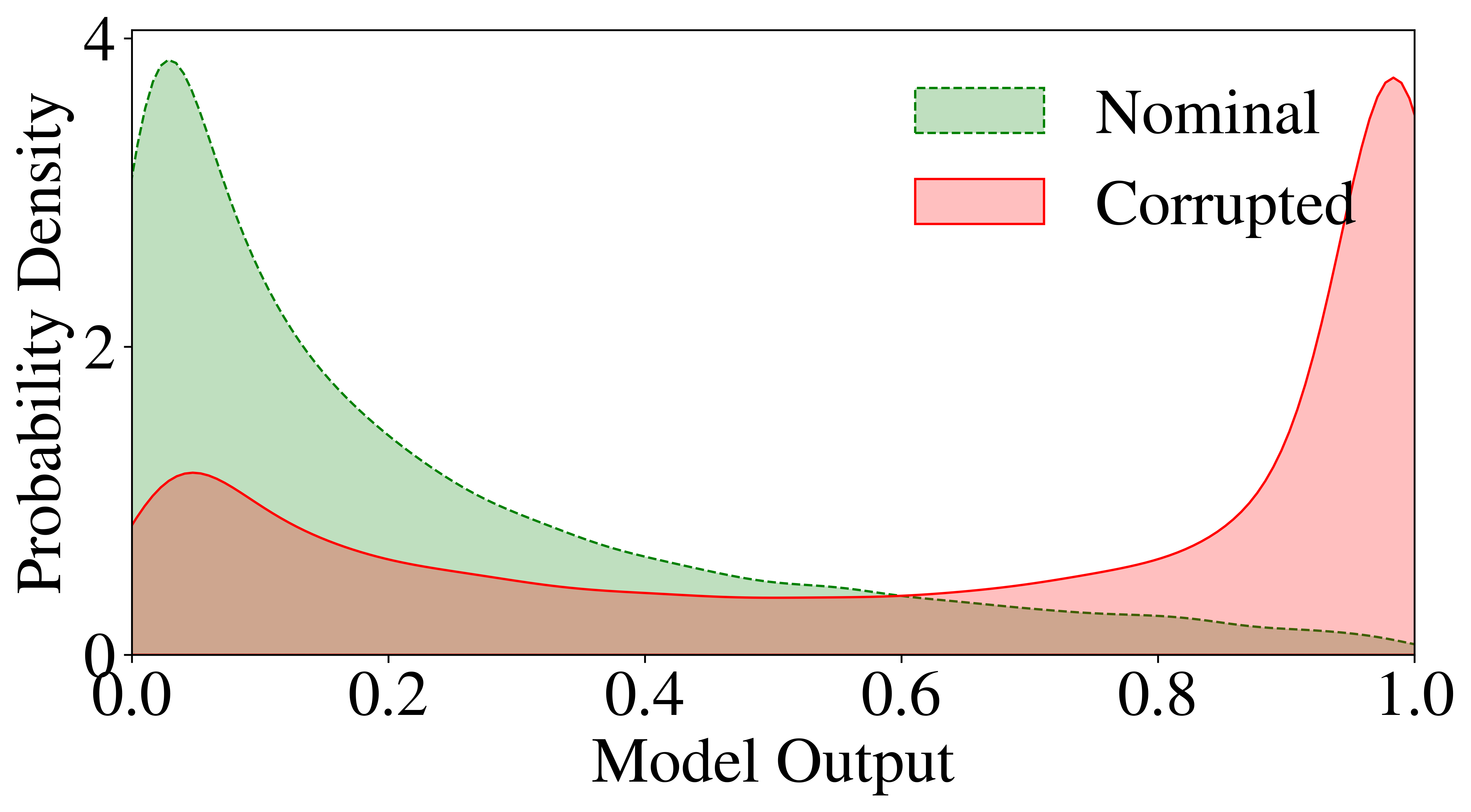}}\qquad
      \subfigure[][b]{\includegraphics[width=0.33\textwidth]{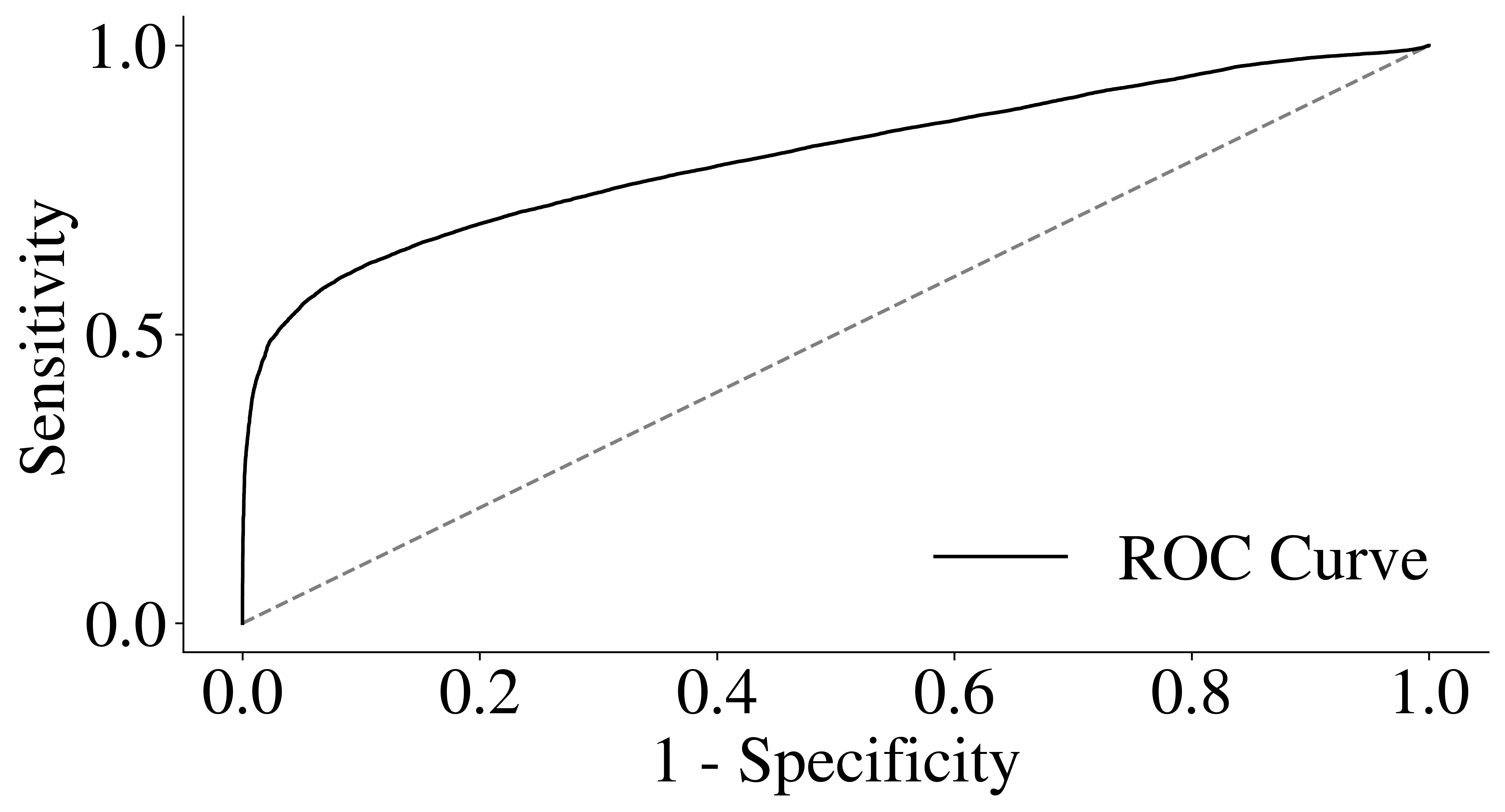}}\qquad
      {\subfigure[][b]{\label{table:discriminator-metrics}\footnotesize
      \begin{tabular}{rc}
        \toprule
        \multicolumn{2}{c}{\textbf{Classifier Metrics}} \\
        \midrule
        Precision & $60.4\%$ \\
        Recall & $85.8\%$ \\
        $F_2$ Score & $79.1\%$ \\
        AUROC & $0.810$ \\
        \bottomrule
         & \\
      \end{tabular}}}
    }
\end{figure*}

\subsection{Learning Multiple Anatomies}
Using the same model architecture as described in \sectionref{section:var-acceleration-methods}, we proposed a modified loss function that aims to maximize the SSIM subject to an EWC regularization term described by \citet{catastrophic-forgetting}
\begin{equation}
\begin{split}
  \mathcal{L}(\tilde{\mathbf{x}}, &\mathbf{x}; \{\theta_i\})=1 - \textnormal{SSIM}(\tilde{\mathbf{x}}, \mathbf{x})\\
  &+\frac{\lambda}{2}\sum_{\textnormal{params}}F_{i}(\theta_i-\theta_{\textnormal{knee, } i})^2 \label{eq:ewc-loss}
\end{split}
\end{equation}
Here, $\{\theta_i\}$ are the trainable model parameters and $\{\theta_{\textnormal{knee}, i}\}$ are the final model parameters after training the model on the fastMRI knee dataset. $F_i$ is the $i$th diagonal element of the Fisher Information Matrix (FIM) of the neural network. $\lambda$ is a constant that dictates the relative importance of the EWC regularization term. Model performance was assessed using the SSIM and PSNR metrics on the holdout test set.

\subsection{Phantom Pre-Training}
To assess the feasibility of transfer learning for accelerated MRI reconstruction, we pre-trained a brain multi-coil E2E-VarNet reconstructor on $70,748$ simulated modified Shepp-Logan phantom slices with the component ellipses randomly distorted with each slice \citep{shepplogan}, taking inspiration from data augmentation techniques described by \citet{mraugment} and others. We then fine-tuned our model for $50$ additional epochs on `fastMRI Mini,' a dataset of $444$ training slices from the fastMRI multi-coil brain training dataset partition, comprising only about $9$ GB and $0.6\%$ of the total available fastMRI multi-coil training brain slices. Our model was trained to maximize the SSIM between the predicted image reconstruction and the ground truth, and the SSIM, NMSE, and PSNR metrics were used for quantitative model evaluation.

\begin{table*}[tbp]
  \floatconts
    {table:reconstructor}
    {\caption{Reconstruction results on the fastMRI knee multi-coil dataset. Statistics reported as either (mean $\pm$ SEM) or (median). Representative image reconstructions are provided in \sectionref{supp:reconstructor}.}}
    {\small
    \begin{tabular}{rccc}
      \toprule
      & SSIM ($\times 10^2$) & NMSE ($\times 10^3$) & PSNR \\
      \cline{2-4} \\[-7px]
      Reconstructor & \multicolumn{3}{c}{\textbf{Variable Acceleration Factor}}\\
      \midrule
      Zero-Filled & $81.7 \pm 0.2$ & $25.1$ & $32.3\pm 0.1$ \\
      Compressed Sensing & $71.3\pm 0.4$ & $22.3$ & $31.7\pm 0.1$ \\
      Sriram et al VarNet & $92.9 \pm 0.1$ & $5.01$ & $38.7\pm 0.1$ \\
      Our VarNet & $\mathbf{94.4 \pm 0.1}$ & $\mathbf{2.61}$ & $\mathbf{42.8\pm 0.1}$ \\
      \midrule
      Reconstructor & \multicolumn{3}{c}{\textbf{Fixed 4x/8x Acceleration Factor}} \\
      \midrule
      Zero-Filled & $70.9 \pm 0.2 / 62.4 \pm 0.2$ & $56.5 / 76.5$ & $27.0\pm 0.1 / 25.6\pm 0.1$ \\
      Compressed Sensing & $71.2 \pm 0.4 / 65.9 \pm 0.4$ & $20.2 / 33.3$ & $31.8 \pm 0.1 / 29.9 \pm 0.1$ \\
      Sriram et al VarNet & $\mathbf{90.0 \pm 0.1} / 87.1 \pm 0.1$ & $\mathbf{5.96} / \mathbf{9.43}$ & $\mathbf{37.5\pm 0.1} / \mathbf{35.8 \pm 0.1}$ \\
      Our VarNet & $\mathbf{90.0 \pm 0.1} / \mathbf{87.2\pm 0.1}$ & $6.48 / 11.4$ & $37.4\pm 0.1 / 35.7\pm 0.1$ \\
      \bottomrule
    \end{tabular}
    }
\end{table*}

\section{Results}
\label{section:results}
\subsection{Detecting Signal Corruption}
\label{section:discriminator-module}
Our discriminator model was able to achieve apparent separation of artifact-corrupted and nominal lines of acquired $k$-space (\figureref{fig:discriminator-graphs}). Following \citet{bydder2002} and others, our model was adapted to process signal in real time during the data acquisition process. This means that any lines flagged as corrupt by our discriminator can still be reacquired during the same scan session.

For this reason, we chose to set a threshold value of $y_{\textnormal{min}}=0.1$ based on the distribution of model output values (\figureref{fig:detector-heatmap}), cognizant of the fact that nominal data inappropriately flagged by our discriminator can still be reacquired, and prioritized a higher classifier recall score over precision (\figureref{table:discriminator-metrics}). This allowed us to capture a larger proportion of the corrupted data. Compared to a non-learning baseline detector described in \sectionref{section:detector-baseline}, our results suggest that our signal corruption detector offers potential clinical utility in MR signal preprocessing applications.

In our hands, GPU-accelerated inference time for real-time artifact detection using our final trained model averaged about 167 msec per $k$-space line. In MR imaging sequences, the time that it takes to for a single $k$-space line acquisition is limited by the \textit{Repetition Time} (TR), which is often between hundreds and thousands of milliseconds \citep{rsna-mr-review}. We therefore expect that real-time artifact detection will not be prohibitively slow in many standard MR imaging applications.

In summary, we offer a ConvNet model for on-the-fly signal processing during accelerated MRI data acquisition and image reconstruction. Its small memory footprint encourages tractable integration within existing scanner and/or cloud-based environments. We encourage future work to (1) investigate applications to more clinically relevant multi-coil data with more sophisticated architectures; (2) find ways to more intelligently integrate the information encoded in corrupted data with subsequent acquisitions, as opposed to simply throwing away flagged data; and (3) validate that the acquired reference ACS lines are indeed nominally uncorrupted through self-consistency or other self-supervised methods.

\subsection{Acceleration Factor Balancing}
\label{section:var-acceleration}

Training our model on variably accelerated data improved its reconstructor performance on input data with acceleration factors that were either greater than 8x or less than 4x when compared to the original E2E-VarNet model \citep{sriram}. However, we also noticed that our model performed comparable to Sriram et al's original E2E-VarNet at 4x acceleration factor, and better than the original model at 8x acceleration factor (\tableref{table:reconstructor}). This suggests that training accelerated MRI reconstruction models on variably accelerated data may offer improved performance at almost all stages of a clinical patient scan.

\subsection{Learning Multiple Anatomies}
\label{section:ewc}
Our results suggest that the EWC regularization penalty term in \equationref{eq:ewc-loss} decreases catastrophic forgetting, albeit by concurrently restricting learning of the second task (reconstructing sparsely sampled brain anatomies) when compared to learning without EWC loss contribution (\tableref{table:ewc}). Similar tradeoffs are reported by \citet{glioma-ewc} and others in sequential learning.

\begin{table*}[t]
  \floatconts
    {table:ewc}
    {\caption{Learning to reconstruct multiple anatomies in accelerated MRI using variably accelerated training data. As in \equationref{eq:ewc-loss}, a value of $\lambda=0$ corresponds to no EWC regularization in learning brain reconstruction after learning knee reconstruction, while a value of $\lambda=+\infty$ corresponds to no brain-specific training. Mean test statistics reported, all SEM values less than $0.1$. Sample image reconstructions from models trained using different values of $\lambda$ are included in \sectionref{section:ewc-supp}.}}
    {\small
    \begin{tabular}{rccccccccccc}
      & \multicolumn{4}{c}{\textbf{Variable Acceleration Factor}} & & \multicolumn{4}{c}{\textbf{8x Acceleration Factor}} \\
      \toprule
      & \multicolumn{2}{c}{\textbf{Brain (Task 2)}} & \multicolumn{2}{c}{\textbf{Knee (Task 1)}} & & \multicolumn{2}{c}{\textbf{Brain (Task 2)}} & \multicolumn{2}{c}{\textbf{Knee (Task 1)}} \\
      \midrule
      $\lambda$ & SSIM & PSNR & SSIM & PSNR & & SSIM & PSNR & SSIM & PSNR \\
      \midrule
      $0$ & $96.7$ & $43.2$ & $93.6$ & $41.4$ & & $91.6$ & $34.7$ & $84.2$ & $34.2$\\
      $3\times 10^2$ & $95.6$ & $41.3$ & $94.2$ & $42.3$ & & $88.1$ & $32.2$ & $86.1$ & $34.9$\\
      $3\times 10^3$ & $95.5$ & $41.0$ & $94.3$ & $42.3$ & & $87.4$ & $31.9$ & $86.8$ & $35.3$\\
      $3\times 10^4$ & $95.0$ & $40.8$ & $94.4$ & $42.6$ & & $86.3$ & $31.4$ & $87.0$ & $35.4$\\
      $3\times 10^5$ & $94.7$ & $40.6$ & $94.6$ & $42.9$ & & $85.8$ & $31.4$ & $87.1$ & $35.7$\\
      $\infty$ & $94.5$ & $40.5$ & $94.6$ & $43.0$ & & $85.5$ & $31.3 $ & $87.2$ & $35.7$\\
      \bottomrule
    \end{tabular}
    }
\end{table*}

\begin{table*}[ht]
  \floatconts
    {table:fisher-overlap}
    {\caption{Fisher overlap $\Omega$ by E2E-VarNet component. $\Omega$ values all multiplied by $100$.}}
    {\small
    \begin{tabular}{rcccccc}
      \toprule
      & \multicolumn{6}{c}{$\lambda$ (EWC Regularization Weighting)}\\
      \cline{2-7}\\[-7px]
      E2E-VarNet Component & $0$ & $3\times 10^2$ & $3\times 10^3$ & $3\times 10^4$ & $3\times 10^5$ & $+\infty$ \\
      \midrule
      Data Consistency & $99.9$ & $99.8$ & $99.2$ & $95.1$ & $94.7$ & $94.3$\\
      Refinement & $95.4$ & $98.7$ & $99.0$ & $91.5$ & $92.6$ & $91.5$\\
      \midrule
      Entire Model & $91.7$ & $95.5$ & $98.2$ & $87.2$ & $82.1$ & $81.9$ \\
      \bottomrule
    \end{tabular}
    }
\end{table*}

To further interrogate our model's ability in learning to reconstruct multiple anatomies, we wanted to explore whether our joint knee-brain reconstructor used similar sets of weights in reconstructing brain and knee anatomies. We computed the Fisher overlap $\Omega$ between brain and knee reconstruction tasks, which is given by
\begin{equation}
  \Omega = 1 - \frac{1}{2}\textnormal{tr}[\hat{F}_A + \hat{F}_B - 2(\hat{F}_A\hat{F}_B)^{1/2}] \label{eq:fisher-overlap}
\end{equation}
where $\hat{F}_A$ and $\hat{F}_B$ are the Fisher Information Matrices of tasks $A$ (knee reconstruction) and $B$ (brain reconstruction), respectively, normalized to unit trace \citep{catastrophic-forgetting}. Here, $\Omega$ is bounded between $0$ and $1$: a value of $\Omega=0$ indicates that the two tasks rely on non-overlapping sets of weights.

In \tableref{table:fisher-overlap}, we report Fisher overlap values as a function of $\lambda$, the weighting on the EWC-derived term as in Equation \ref{eq:ewc-loss}. We found that models trained with EWC regularization use more similar sets of weights in reconstructing sparsely sampled brain and knee datasets (i.e. $\lambda=3\times 10^2, 3\times 10^3$) than a model that has forgotten previously learned knee anatomies ($\lambda=0$), or a model that that has never seen a brain anatomy ($\lambda=+\infty$). However, as the plasticity of the model weights decreases, corresponding to larger values of $\lambda$ (i.e. $\lambda=3\times 10^4$ and $\lambda=3\times 10^5$), the model begins to use increasingly different sets of weights, which also corresponds to decreasing model performance (\tableref{table:ewc}).

\begin{table*}[htp]
  \floatconts
    {table:transferlearning}
    {\caption{Transfer learning for multi-coil brain reconstruction with fixed 4x and 8x acceleration factors. Test statistics reported as (mean $\pm$ SEM).}}
    {\small
    \begin{tabular}{ccccc}
      \toprule
      Pre-Training & Fine-Tuning & SSIM ($\times 10^2$) & NMSE ($\times 10^3$) & PSNR \\
      \midrule
      --- & fastMRI Mini & $86.9 \pm 0.04$ & $23.7\pm 0.1$ & $32.2\pm 0.03$ \\
      fastMRI Mini & fastMRI Mini & $87.4\pm 0.04$ & $22.1\pm 0.1$ & $32.6\pm 0.03$ \\
      Simulated Phantoms & fastMRI Mini & $88.1 \pm 0.04$ & $22.2\pm 0.2$ & $32.5\pm 0.03$ \\
      fastMRI & fastMRI Mini & $89.7 \pm 0.04$ & $16.9\pm 0.1$ & $33.5\pm 0.03$ \\
      \midrule
      \multicolumn{2}{c}{Full fastMRI dataset for 50 epochs} & $93.3\pm 0.02$ & $8.66\pm 0.05$ & $36.4 \pm 0.02$\\
      \bottomrule
    \end{tabular}
    }
\end{table*}

We further stratified the Fisher overlap by the component type within the E2E-VarNet reconstructor model. Described in detail by \citet{sriram}, the E2E-VarNet model features two general components in each iterative step: (1) a \textit{data consistency} module that ensures the predicted $k$-space model output is consistent with the acquired lines; and (2) a \textit{refinement} module that resolves the details of the anatomy in image space. We found that the Fisher overlap is consistently lower in the refinement module than in the data consistency module. This may be explained by a greater difference in the brain and knee datasets in image space than in Fourier space.

\subsection{Phantom Pre-Training}
\label{section:tl}
Pre-training an E2E-VarNet reconstructor using a simulated Shepp-Logan phantoms improves model performance in reconstructing clinically acquired brain images when compared to working solely with a restricted dataset, based on both quantitative (\tableref{table:transferlearning}, \figureref{fig:tl}) and qualitative (\sectionref{section:tl-supp}) results. Of note, pre-training took less than a day and fine-tuning under 3 hours using a single 16 GB Tesla P100 GPU. These metrics allow for rapid, inexpensive model iteration and experimentation before final benchmarking on a more sophisticated compute cluster.

Although MRI brain data is widely available through the fastMRI dataset among others, consolidated MR datasets of other clinically relevant anatomies, such as the abdomen and spine, are limited. In these cases, model pre-training using appropriately simulated phantom data may prove useful in experimenting with models to reconstruct largely unexplored anatomies.

\section{Discussion and Future Work}
\label{section:discussion}
Together with work on intelligent MR data acquisition \citep{bakker, yin2021end, greedy, mri-active-acq-zhang, loupe} and other clinically relevant technologies \citep{yaman2022zeroshot}, our results serve as a springboard towards clinical adaptation of learning-based accelerated MRI reconstruction. We investigated potential improvements to data pre-processing pipelines, such as through identifying contaminated MR signal data and using variable acceleration factors during model training, to make image reconstructors more robust to a variety of simulated clinical scenarios. Furthermore, we explored the ability of MR reconstructor models to reconstruct different anatomies, and demonstrated that model pre-training using simulated phantom data could prove useful in developing future learning-based methods.
 
Further work remains to be done for clinical adaptation of accelerated MRI reconstruction. For example, \citet{yolo} reported that optimizing the structural similarity metric during training of image reconstructors can result in image smoothing that reduces the detectability of finer lesions. Future model training may revolve around the recent availability of fastMRI+, which offers clinical annotations of lesions within the fastMRI dataset \citep{fastmriplus}.

Finally, alternatives to Cartesian $k$-space acquisition trajectories may also be explored. Much of the current work involving the fastMRI dataset has assumed a Cartesian sampling pattern, which can result in both coherent and incoherent artifacts in the final image reconstruction \citep{yaman2022zeroshot}, in addition to non-physical gradient slew rates that could lead to significant data corruption if not appropriately accounted for.

\section*{Code Availability}
\label{section:code-availability}
The code for this project is available at \href{https://github.com/michael-s-yao/accMRI}{github.com/michael-s-yao/accMRI} and is licensed under the MIT License. Portions of our codebase were adapted from the \texttt{facebookresearc} \texttt{h/fastMRI} Github repository at \href{https://github.com/facebookresearch/fastMRI}{github.com/facebookresearch/fastMRI} --- we thank the contributors and maintainers of this project for making it accessible for academic use.

\section*{Acknowledgements}
The authors thank John Stairs, Ajay Paidi, Kyle Greenslade, and Himanshu Sahoo at Microsoft Research, and Dylan Tisdall at the University of Pennsylvania, for their helpful discussions. We also sincerely thank Desney Tan at Microsoft Research for project support.

Data used in the preparation of this manuscript were obtained from the NYU fastMRI Initiative database at \href{https://fastmri.med.nyu.edu}{fastmri.med.nyu.edu} \citep{zbontar2018fastMRI}. As such, NYU fastMRI investigators provided data but did not participate in analysis or writing of this manuscript. A listing of NYU fastMRI investigators, subject to updates, can be found at \href{https://fastmri.med.nyu.edu}{fastmri.med.nyu.e} \href{https://fastmri.med.nyu.edu}{du}. The primary goal of fastMRI is to test whether machine learning can aid in the reconstruction of medical images.

\section*{Disclosure of Funding}
This research was supported by Microsoft Research and its affiliates. M.S.Y. is supported by the Microsoft Research Internship Program, as well as the NIH Medical Scientist Training Program T32 GM007170 and the NIH Biomedical Imaging Training Program T32 EB009384 at the University of Pennsylvania.

\section*{Author Contributions}
M.S.Y. and M.S.H. conceived the study. M.S.Y. planned and performed experiments, and processed and analyzed the data before preparing the manuscript. M.S.H. supervised and coordinated the research and helped revise the manuscript.

\section*{Competing Interests}
The authors declare no competing interests related to this work.

{\small
\bibliography{yao22}
}

\appendix
\section{Implementation Details}
\label{section:implementation-details}

\subsection{Detecting Signal Corruption}
\label{section:discriminator-implementation-details}
Our model was trained on the fastMRI single-coil knee training dataset and validated on half of the fastMRI single-coil knee validation dataset. Final model testing reported herein in \figureref{fig:discriminator-graphs} was performed on the holdout remaining half of the fastMRI validation dataset consisting of $N=3567$ test slices.

To simulate artifacts encountered in the clinical setting, we modeled in-plane affine rotational motion around the isocenter between 20 and 50 degrees, and both point and line spikes in $k$-space where the magnitude of random selected point(s) are set to a random value uniformly chosen between 1- and 2- times the maximum magnitude in the slice dataset. Additional details of our implementation are included in our source code in \sectionref{section:code-availability}.

\subsection{Acceleration Factor Balancing}
\label{section:var-acceleration:implementation-details}
To simulate variable acceleration factors, we randomly masked the slices in the fastMRI training by sampling the number of unmasked lines from a uniform distribution $\mathcal{U}\left[\texttt{min\_lines}, \texttt{width}\right]$, where $\texttt{min\_lines}=16$ and $\texttt{width}$ is the number of Cartesian $k$-space columns in the slice. The acceleration factor could then vary anywhere between approximately 20x to a fully sampled dataset. Because the acceleration factor is proportional to the inverse of the fraction of unmasked lines, the distribution of acceleration factors in our training dataset therefore followed an inverse power law. Training was conducted on the fastMRI multi-coil knee training dataset partition. The results reported in \tableref{table:reconstructor} were generated from the fastMRI multi-coil knee validation dataset, which was withheld during training of our model.

\subsection{Learning Multiple Anatomies}
Similar to \sectionref{section:var-acceleration:implementation-details}, we once again trained an E2E-VarNet with 8 cascade layers and similarly simulated variable acceleration factors as described previously. We trained our model for $100$ epochs using an Adam optimizer. In the first $50$ epochs, our model learned from the fastMRI multi-coil \textit{knee} dataset partition, while in the last $50$ epochs our model was only trained on the fastMRI multi-coil \textit{brain} dataset.

Models were trained on the fastMRI multi-coil knee and multi-coil brain training dataset partitions. For model validation, we selected a small subsample of the fastMRI knee and brain multi-coil validation datasets consisting of $296$ knee slices and $128$ brain slices. Model testing was performed on the heldout fastMRI knee and brain multi-coil validation dataset partitions excluding the slices used for model validation.

During training, the EWC penalty term in \equationref{eq:ewc-loss} required the diagonal elements $\{F_i\}$ of the Fisher Information Matrix of our VarNet only trained on knee anatomies (Task 1), which was already trained from our previously described work (\sectionref{section:var-acceleration}). For computational tractability, we chose to calculate $\{F_i\}$ using a smaller subsample of the fastMRI knee multi-coil training dataset partition constituting a total of $N=296$ slices from $8$ different volumes. To calculate the Fisher Information Matrices for determining Fisher Overlap, we used the model validation datasets described above.

\subsection{Phantom Pre-Training}
\label{section:tl:implementation-details}
In simulating our Shepp-Logan phantom datasets for pre-training, we applied random affine transformations to constitutive ellipses composing each three-dimensional Shepp-Logan phantom dataset, and applied global phase rolling to add complex phase to the images. We also applied Gaussian blurring with standard deviation of $\sigma=1$ to the image, followed by uncorrelated noise and receiver coil modeling to obtain a simulated multi-coil $k$-space dataset from a real Shepp-Logan phantom slice.

For our transfer learning experiments, all input signal data for model training and testing was fixed at 4x and 8x acceleration factors. We tested our model on the fastMRI multi-coil validation dataset partition, excluding the volumes that were used in the fastMRI Mini dataset for fine-tuning.

\begin{figure*}[ht]
    \floatconts
      {fig:discriminator-images}
      {\caption{Sample reconstructions using the baseline single-coil U-Net fastMRI model with and without signal discriminator preprocessing. Lines that were flagged for removal were removed from the slice dataset prior to image reconstruction.}}
      {\includegraphics[width=\textwidth]{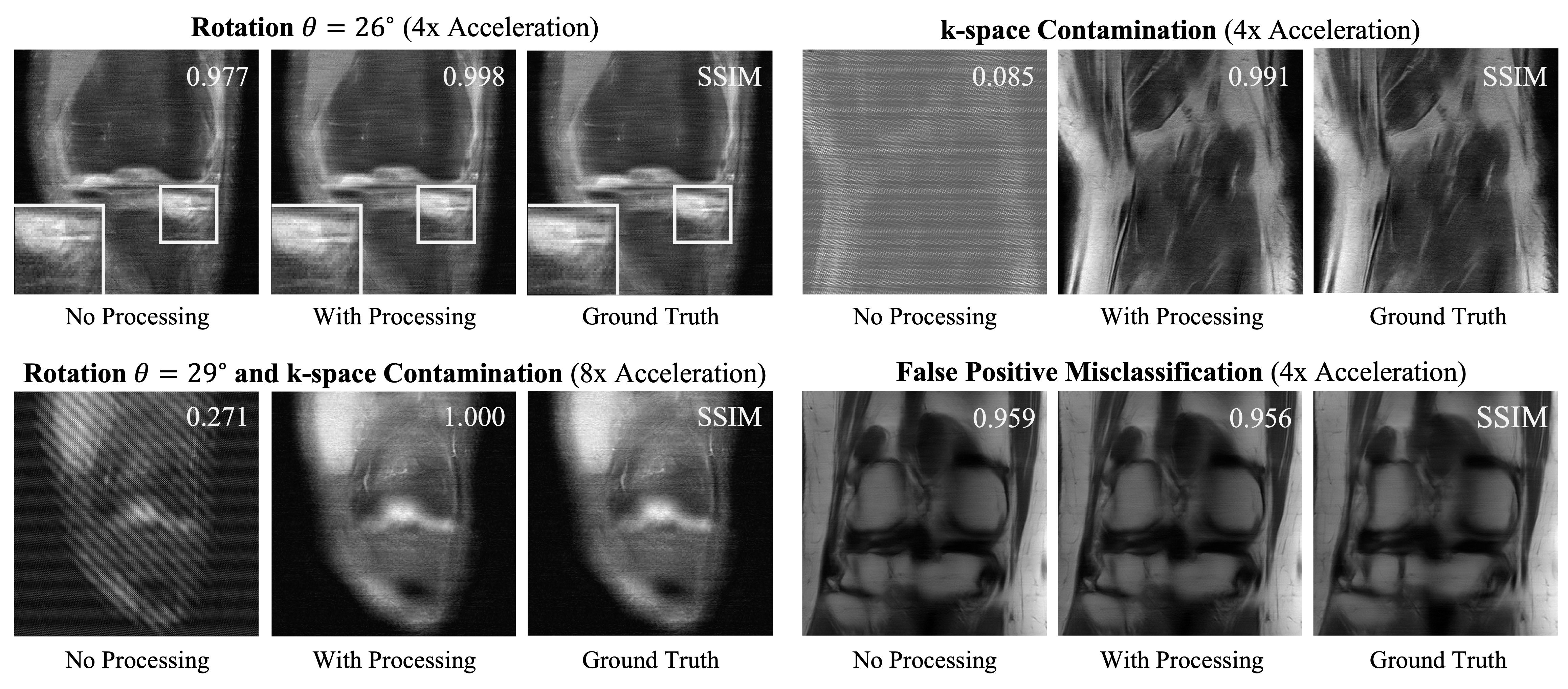}}
\end{figure*}

\section{Additional Results}
\subsection{Detecting Signal Corruption}
In this section, we provide additional discussion on the motivation for, and limitations of, our learning-based approach to estimating the fidelity of acquired $k$-space lines in accelerated MRI applications. First, we discuss the motivation behind our ConvNet architecture in \sectionref{section:discriminator-module}, followed by comparison with a non-learning baseline algorithm and limitations of our proposed method.

\begin{figure*}[t]
  \floatconts
    {fig:covar}
    {\caption{Sample $k$-space line auto-correlation maps plotting the element-wise magnitude (top) and phase (bottom) of $\mathbf{R}_{\mathbf{XX}}$ from (\ref{eq:autocorr}). The red hollow arrow points to the particular line (ie column index $0$, $21$, $64$, and $97$) with which the correlation map was generated. The black solid arrow points to the center of $k$-space. All plotted values for the magnitude plots scaled by $10^{10}$.}}
    {
    \includegraphics[width=\textwidth]{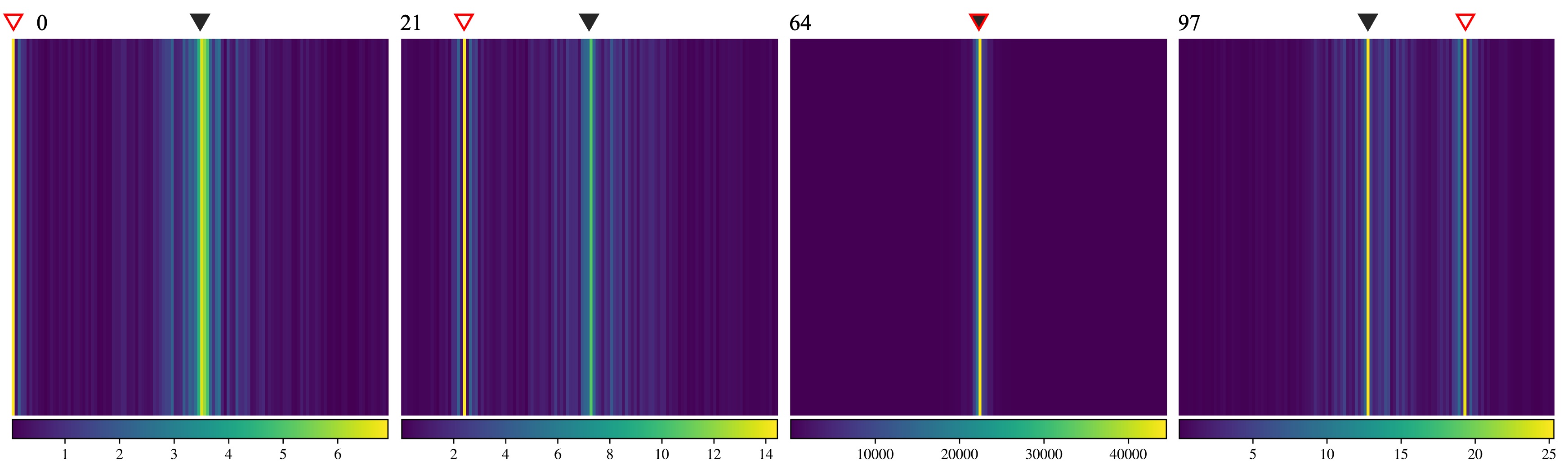}
    \includegraphics[width=\textwidth]{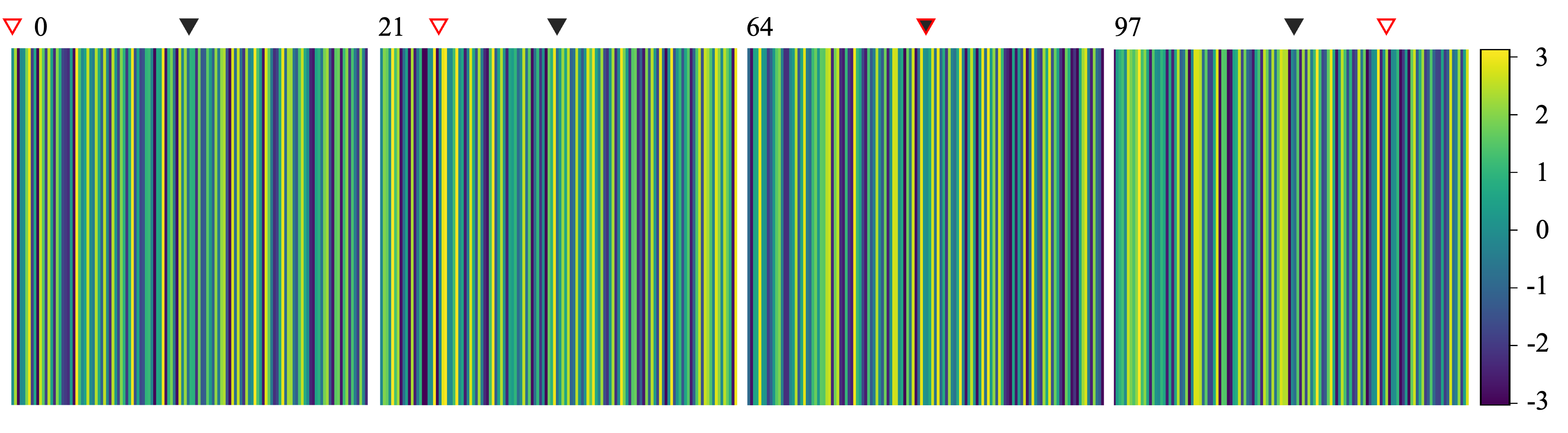}
    }
\end{figure*}

\paragraph{Representative Images}
\label{section:discriminator-qualitative}
Using the fastMRI baseline single-coil knee U-Net reconstructor from the open-source \href{https://github.com/facebookresearch/fastMRI}{\texttt{facebookresearch/fastMRI}} Github repository, we reconstructed accelerated $k$-space data both with and without signal pre-processing using our trained model (\figureref{fig:discriminator-images}). Our neural network successfully identified both rigid rotational motion and lines contaminated with $k$-space spiking. Removing the lines flagged by our model resulted in improvements to the structural similarity metric (SSIM) when compared to the ground truth, which is the U-Net reconstruction with all of the simulated corrupted lines removed. In cases of false positive misclassification, the SSIM decreased slightly, albeit with minimal qualitative reduction in image quality, due to having less data available at reconstruction time.

\paragraph{Estimating the Correlation of \texorpdfstring{$k$}{k}-space Lines}
\label{section:kspace-correlation}
To inform the design of our discriminator network architecture, we first sought to estimate the correlation between different pairs of $k$-space lines. We hypothesized that observed Fourier data with higher cross-correlation shared with newly acquired $k$-space lines would be the most useful in predicting the fidelity of the new Fourier data.

We first took the $N=7135$ slices from the fully sampled fastMRI single-coil knee validation dataset partition and center-cropped the slices to $128\times 128$ before flattening them to a vector of size $M=128^2$. These slices could then be used to construct the auto-correlation matrix
\begin{equation}
    \mathbf{R}_{\mathbf{X}\mathbf{X}}=(N-1)^{-1}\mathbf{X}\mathbf{X}^H \label{eq:autocorr}
\end{equation}
where $\mathbf{X}\in\mathbb{C}^{M\times N}$ is the complex feature matrix and $\mathbf{X}^H$ is the Hermitian transpose of $\mathbf{X}$. Because we assumed a Cartesian $k$-space sampling pattern, we summed over the readout dimension to obtain a proxy for the correlation between lines, as opposed to individual points, in $k$-space. We then plotted the element-wise magnitude (and phase) of $\mathbf{R}_{\mathbf{X}\mathbf{X}}$, which can be thought of as a metric for phase clustering around the mean phase difference (or the mean phase difference) between any two $k$-space lines.

Our results suggested that each Cartesian line shared the greatest correlation with the center fraction of $k$-space (\figureref{fig:covar}), suggesting that comparing newly acquired high-frequency lines with the ACS would be most effective in assessing the fidelity of said data. This motivated the design of our ConvNet discriminator model, which convolves an acquired high-frequency $k$-space line with each line in the ACS to estimate the fidelity of the acquired data (\figureref{figure:cnn-architecture-text}).

\paragraph{Baseline Algorithm Comparison}
\label{section:detector-baseline}

\begin{figure*}[ht]
  \floatconts
    {fig:baseline-discriminator}
    {\caption{Probability density distributions of baseline discriminator outputs (left) and receiver operating characteristic (ROC) curve (right) on test dataset.}}
    {
      \includegraphics[width=0.48\textwidth]{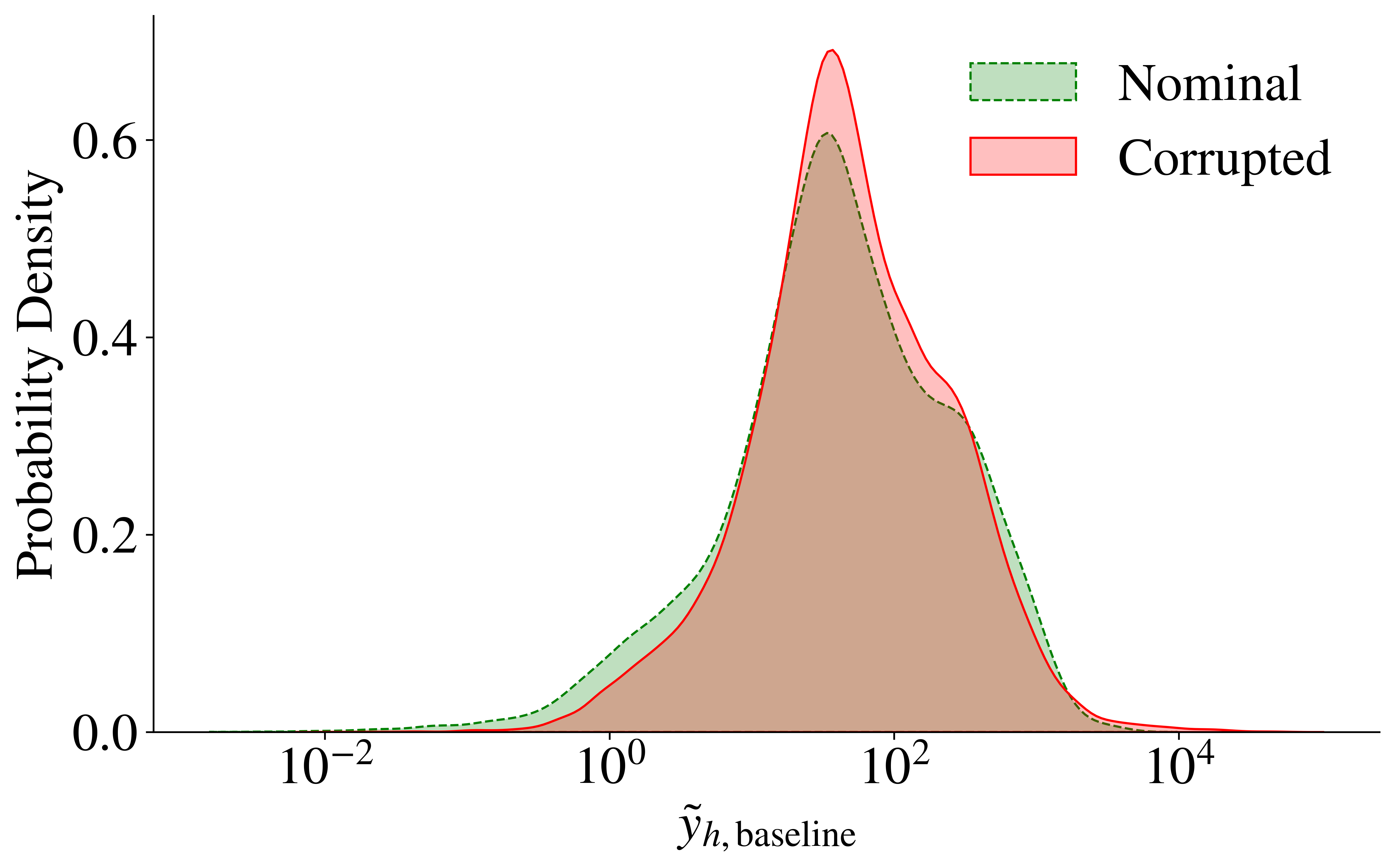}
      \includegraphics[width=0.48\textwidth]{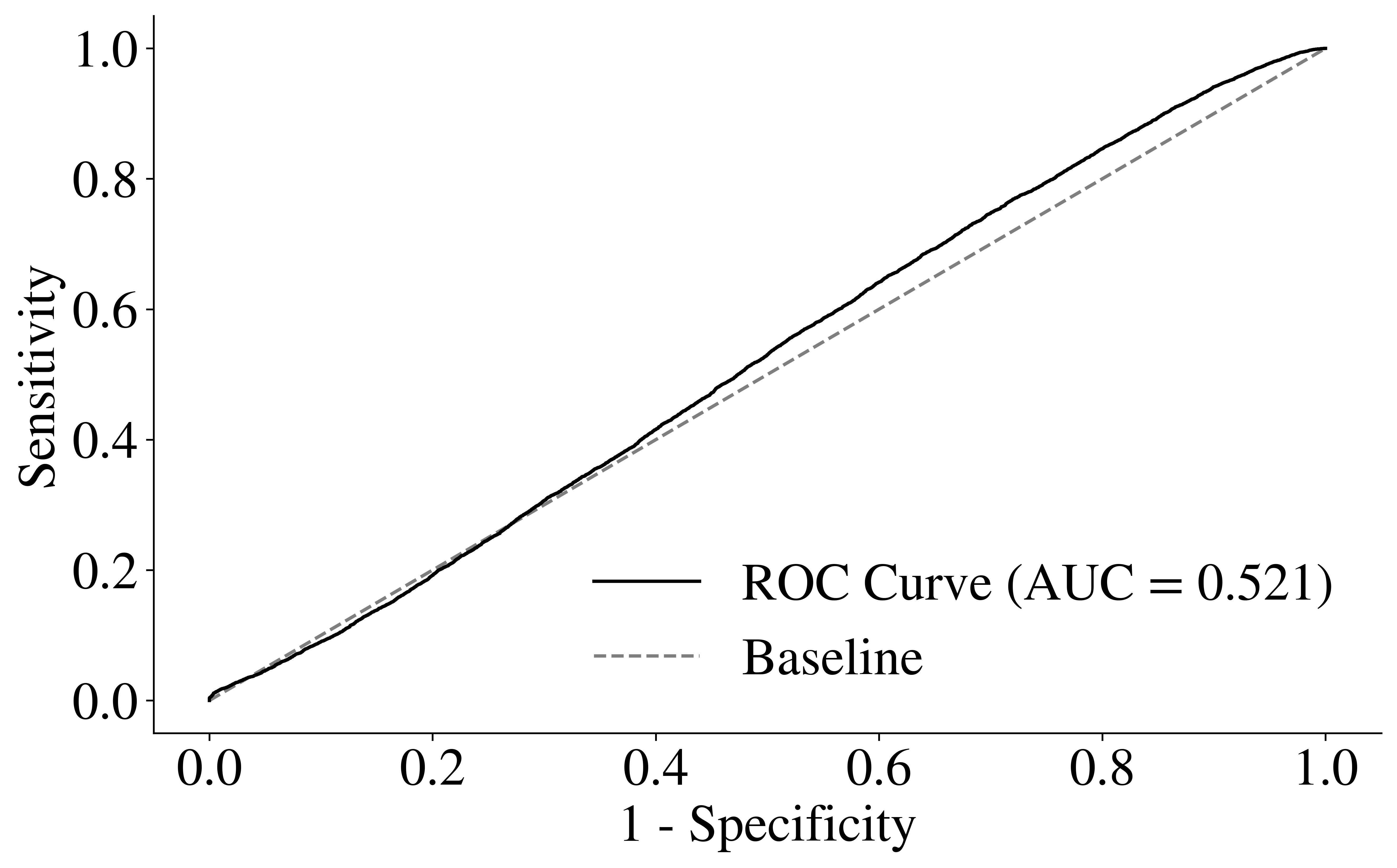}
    }
\end{figure*}

Given our results in \figureref{fig:covar}, we wanted to explore whether learning-based methods were necessary for our task of detecting corrupted $k$-space data. Classical methods are often faster than neural networks at runtime, and can be easier to understand and implement as well.

\begin{table*}[ht]
  \floatconts
    {table:reconstructor-brain-supp}
    {\caption{Reconstruction results on the fastMRI brain multi-coil dataset. Statistics reported as either (mean $\pm$ SEM) or (median).}}
    {\small
    \begin{tabular}{rccc}
      \toprule
      & SSIM ($\times 10^2$) & NMSE ($\times 10^3$) & PSNR \\
      \cline{2-4} \\[-9px]
      Reconstructor & \multicolumn{3}{c}{\textbf{Variable Acceleration Factor}}\\
      \midrule
      Zero-Filled & $84.7 \pm 0.08$ & $22.1$ & $32.9\pm 0.05$ \\
      Compressed Sensing & $70.5\pm 0.06$ & $13.7$ & $33.5\pm 0.08$ \\
      Sriram et al VarNet & $96.4\pm 0.04$ & $3.73$ & $41.5\pm 0.04$ \\
      Our VarNet & $\mathbf{96.7\pm 0.02}$ & $\mathbf{1.55}$ & $\mathbf{43.3\pm 0.03}$ \\
      \midrule
      Reconstructor & \multicolumn{3}{c}{\textbf{Fixed 4x/8x Acceleration Factor}} \\
      \midrule
      Zero-Filled & $74.9 \pm 0.05/67.8 \pm 0.05$ & $56.2/75.1$ & $27.7\pm 0.03/26.4\pm 0.03$ \\
      Compressed Sensing & $71.4\pm 0.06/65.4\pm 0.05$ & $16.0/33.5$ & $32.7\pm 0.1/29.8\pm 0.1$ \\
      Sriram et al VarNet & $94.4\pm 0.02/91.5\pm 0.03$ & $5.28/10.9$ & $39.4\pm0.02/36.0\pm0.03$ \\
      Our VarNet & $\mathbf{94.9\pm 0.02}/\mathbf{91.7\pm 0.03}$ & $\mathbf{3.92}/\mathbf{9.55}$ & $\mathbf{40.4\pm 0.02}/\mathbf{36.9\pm 0.03}$ \\
      \bottomrule
    \end{tabular}
    }
\end{table*}

We hypothesized that an algorithm based on signal cross-correlation between acquired and ACS lines may offer reasonable discriminator performance. Because such detection methods under the constraints of accelerated MRI have not been well-reported in prior literature, we proposed calculating the following quantity $\tilde{y}_{h, \text{baseline}}$ for each acquired high-frequency line $\mathbf{k}_h$:
\begin{equation}
\begin{split}
  &\tilde{y}_{h, \text{baseline}} =\frac{1}{\text{len}(\text{ACS})}\sum_{\mathbf{k}_{\ell}\in\text{ACS}}\\
  &\qquad\frac{\sum_{\text{readout}}\text{abs}(\mathbf{k}_{h}\star \mathbf{k}_{\ell})_{\text{observed}}}{\sum_{\text{readout}}\text{abs}(\mathbf{k}_{h}\star\mathbf{k}_{\ell})_{\text{ground truth}}}
\end{split}
\end{equation}
Here, $\star$ is the cross-correlation operator and $\sum_{\text{readout}}$ represents a summation over the readout dimension. $\text{abs}(\cdot)$ is the element-wise absolute value operator. Unlike our learning-based method introduced in the main text (\sectionref{section:discriminator-module}), the output $\tilde{y}_{h, \text{baseline}}$ is only bounded from below by $0$, and may be arbitrarily large.

\begin{figure*}[ht]
  \floatconts
    {fig:ewc-by-epoch}
    {\caption{(Left) Demonstrating catastrophic forgetting of learned knee image reconstruction after model training on brain data. (Right) EWC rescues the reconstructor from catastrophic forgetting. Data plotted at 8x acceleration factor.}}
    {
    \includegraphics[width=0.48\textwidth]{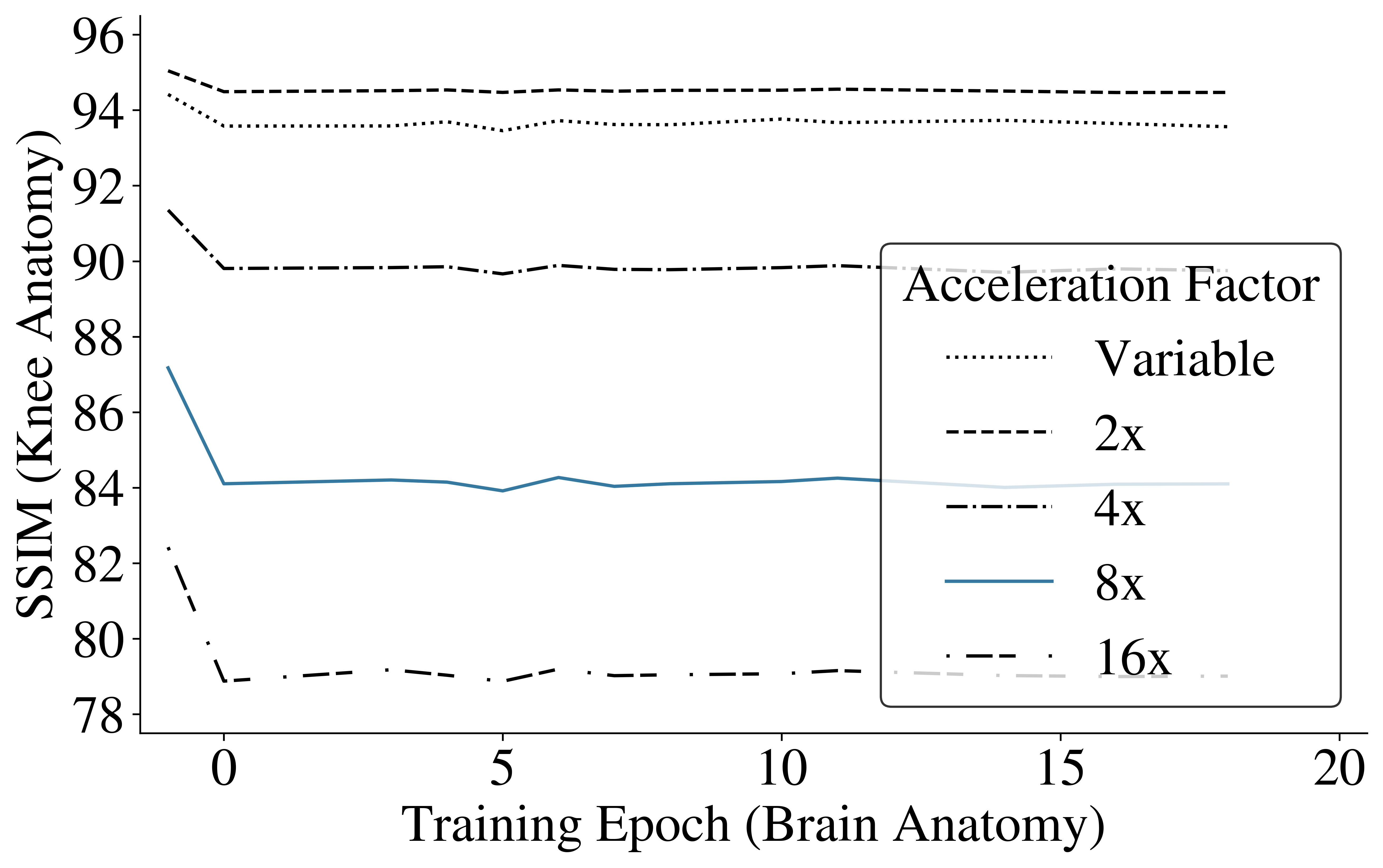}
    \includegraphics[width=0.48\textwidth]{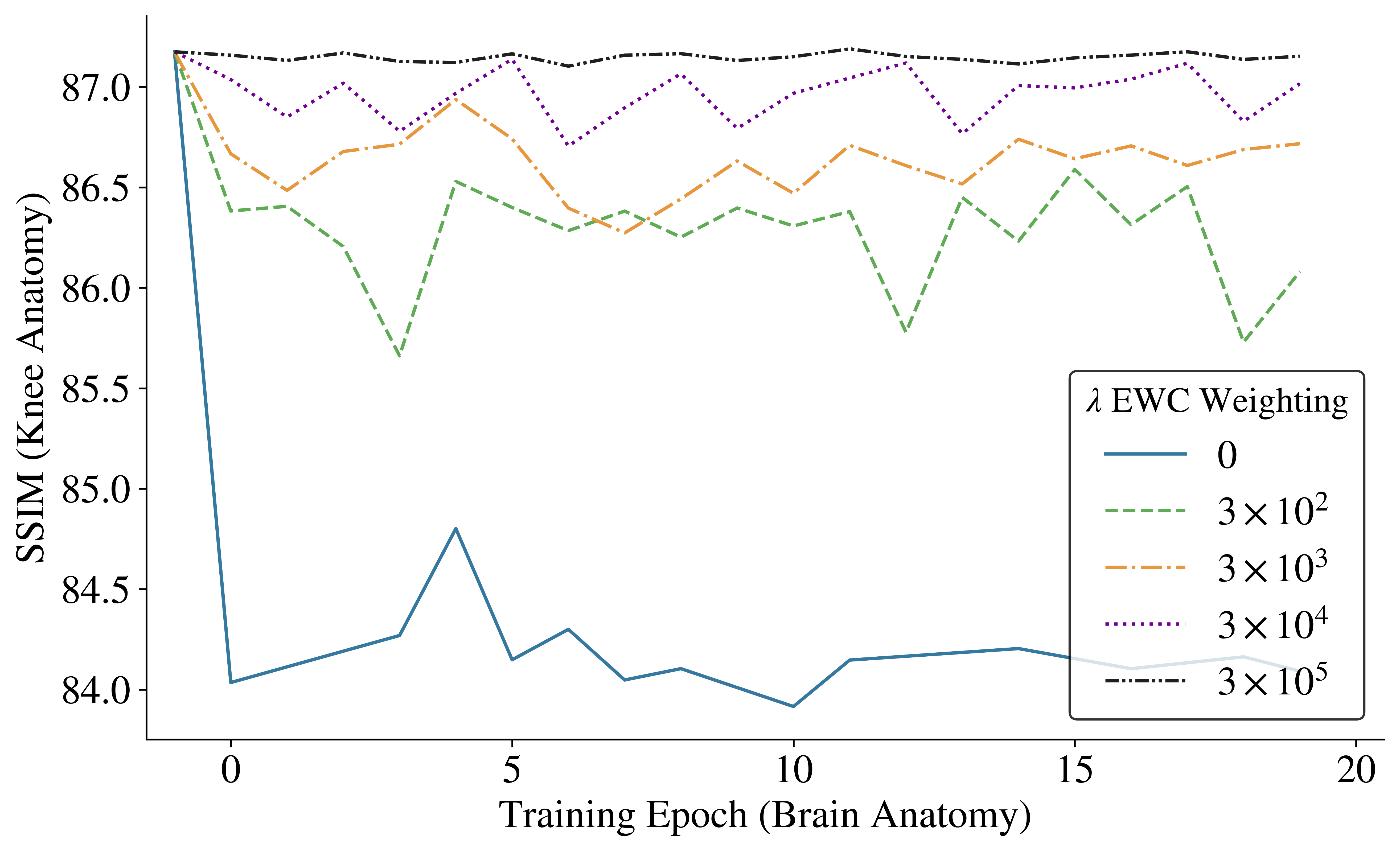}
    }
\end{figure*}

\begin{figure*}[htpb]
  \floatconts
    {fig:ewc-images}
    {\caption{Sample image reconstructions at 8x acceleration factor from VarNet models trained with different EWC weighting as in \equationref{eq:ewc-loss}.}}
    {\includegraphics[width=\textwidth]{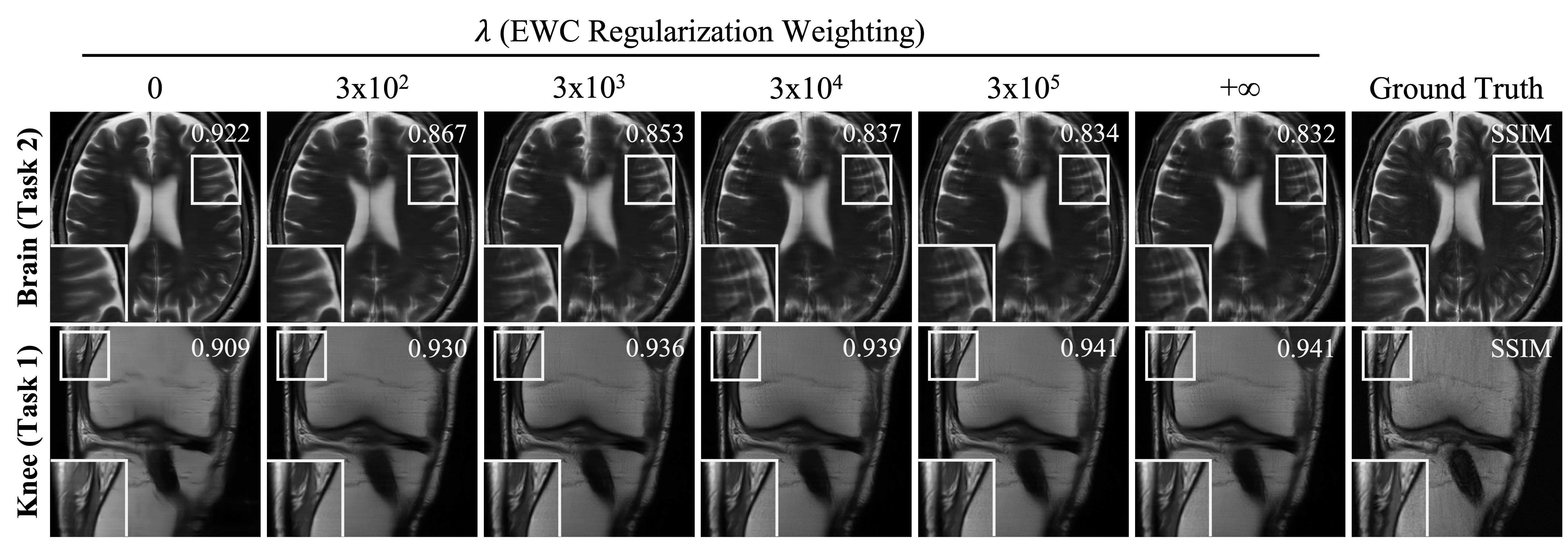}}
\end{figure*}

In \figureref{fig:baseline-discriminator}, we plot the probability density versus $\tilde{y}_{h, \text{baseline}}$ values for both the corrupted and uncorrupted $k$-space lines, simulating both rigid rotational motion and $k$-space contamination as in \sectionref{section:discriminator-implementation-details}. While the plotted distributions are not identical, there is no substantial qualitative difference in the position and shape of the distributions that could be used to discern elements of the two populations meaningfully. Furthermore, the area under the ROC curve (AUC) was $0.521$, which suggests minimal ability to discriminate between nominal and corrupted $k$-space lines. From the threshold values tested, the maximum $F_2$ score we observed using this baseline detector algorithm was $57.0\%$, corresponding to precision and recall values of $51.9\%$ and $58.5\%$ respectively.

Overall, these results suggest that our proposed non-learning-based detector algorithm provides little utility in identifying corrupted data for applications in accelerated MRI, and provides further support in using network networks for this task, as in \sectionref{section:discriminator-module}.

\paragraph{Affine Transforms in Modeling Bulk Patient Motion}
\label{supp:affine-transforms}
In \sectionref{section:discriminator-module}, we proposed a ConvNet-based neural network to detect corrupted lines during a patient scan and flag them for subsequent reacquisition. Our model was informed by our desire to design a $k$-space `discriminator module' for sparsely sampled signal data at any clinically relevant acceleration factor. However, preliminary experiments indicated that our ConvNet model was unable to accurately detect simulated bulk translational motion --- even if both $k_x$ and $k_y$ components were fed into the model. In this section, we offer an explanation for this empirical observation.

We first define $I(x, y)$ as the image of a particular slice. The Fourier transform of the image $\tilde{I}(k_x, k_y)$ is defined as
\begin{equation}
\begin{split}
    &\tilde{I}_{\textnormal{ground truth}}(k_x, k_y)\\
    &\quad=\frac{1}{2\pi}\int_{-\infty}^{+\infty}dx\textnormal{ }e^{-ik_xx}\\
    &\quad\quad\quad\int_{-\infty}^{+\infty}dy\textnormal{ }e^{-ik_yy}I(x, y) \label{eq:ft-definition}
\end{split}
\end{equation}
A bulk translation of the patient in image space by a displacement vector $(\Delta x, \Delta y)$ thus corresponds to a global phase shift in the Fourier transform
\small
\begin{equation}
\begin{split}
    &\tilde{I}_{\textnormal{bulk translation}}(k_x, k_y)\\
    &=\frac{1}{2\pi}\int_{-\infty}^{+\infty}dx\textnormal{ }e^{-ik_xx}\int_{-\infty}^{+\infty}dy\textnormal{ }e^{-ik_yy}\\
    &\qquad\qquad I(x-\Delta x, y-\Delta y)
\end{split}
\end{equation}
\normalsize
We can shift the integration variables to rewrite our expression as
\begin{equation}
\begin{split}
    &\tilde{I}_{\textnormal{bulk translation}}(k_x, k_y)\\
    &=\frac{1}{2\pi}\int_{-\infty}^{+\infty}dx\textnormal{ }e^{-ik_x(x+\Delta x)}\\
    &\qquad\qquad \int_{-\infty}^{+\infty}dy\textnormal{ }e^{-ik_y(y+\Delta y)}I(x, y)
\end{split}
\end{equation}
Factoring and simplifying, we have
\begin{equation}
\begin{split}
    &\tilde{I}_{\textnormal{bulk translation}}(k_x, k_y)\\
    &=\frac{1}{2\pi}e^{-i(k_x\Delta x+k_y\Delta y)}\left[\int_{-\infty}^{+\infty}dx\textnormal{ }e^{-ik_xx}\right.\\
    &\qquad\qquad \left.\int_{-\infty}^{+\infty}dy\textnormal{ }e^{-ik_yy}I(x, y)\right]\\
    &=e^{-i(k_x\Delta x+k_y\Delta y)}\tilde{I}_{\textnormal{ground truth}}(k_x, k_y)\\
    &:=e^{-i\theta(k_x, k_y)}\tilde{I}_{\textnormal{ground truth}}(k_x, k_y)
\end{split}
\end{equation}
when compared to the ground truth in \equationref{eq:ft-definition}. Therefore, in order for a discriminator module to detect bulk translational motion in Fourier space, it needs to be sensitive to the relative phase of an acquired $k$-space line relative to only the center fraction of low-frequency lines, which are assumed to be artifact-free. However, our experiments found that there is little correlation between the \textit{phases} between different $k$-space lines when compared to the correlation between magnitudes (\figureref{fig:covar}). Based on these results, we concluded that our model architecture proposed in \sectionref{section:discriminator-module} for a signal preprocessing module specific for sparsely sampled datasets would be insufficient for identifying artifacts due to bulk translational patient motion. This is because unlike other motion correction algorithms \citep{bydder2002, loktyushin2012, johnson2018} that may differentiate potential phase shifts using both local and global features within a fully sampled Fourier dataset, applications in accelerated MRI cannot assume a fully acquired dataset to work with. Future work building off of our proof-of-concept experiments may involve incorporating scanner navigator data and coil positioning information to better discern potential translational motion.

\begin{figure*}[htpb]
  \floatconts
    {fig:tl}
    {\caption{Representative reconstructions of axial FLAIR (first row) and T2-weighted (second row) brain slices at 4x and 8x acceleration factors, respectively, using varying training pipelines.}}
    {\includegraphics[width=\textwidth]{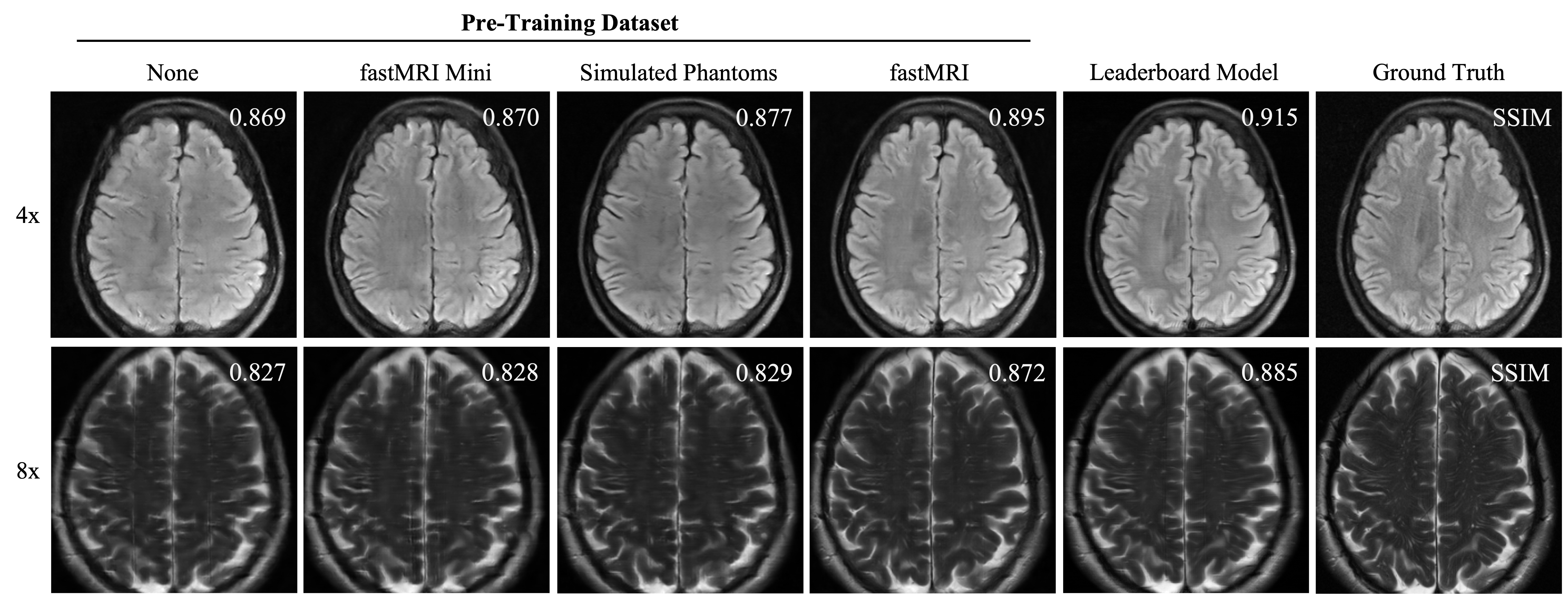}}
\end{figure*}

\begin{figure*}[htpb]
    \floatconts
      {fig:reconstructor-knee}
      {\caption{Representative reconstructions of the same coronal proton density-weighted image of the knee given different acceleration factors (rows) using different reconstruction algorithms (columns).}}
      {\includegraphics[width=\textwidth]{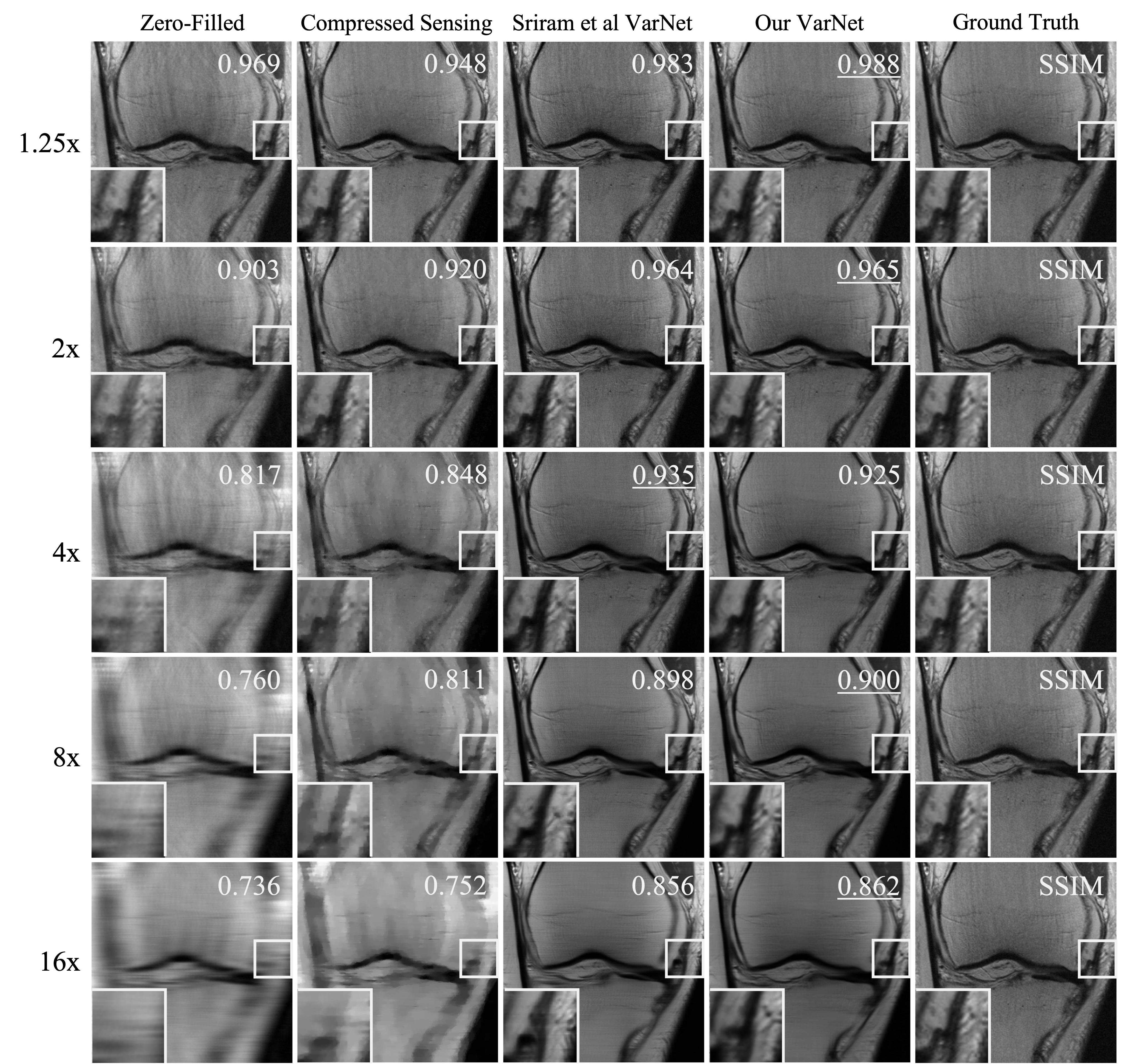}}
\end{figure*}

\begin{figure*}[htpb]
    \floatconts
      {fig:reconstructor-brain}
      {\caption{Representative reconstructions of the same axial T1-weighted slice of the brain given different acceleration factors (rows) using different reconstruction algorithms (columns).}}
      {\includegraphics[width=\textwidth]{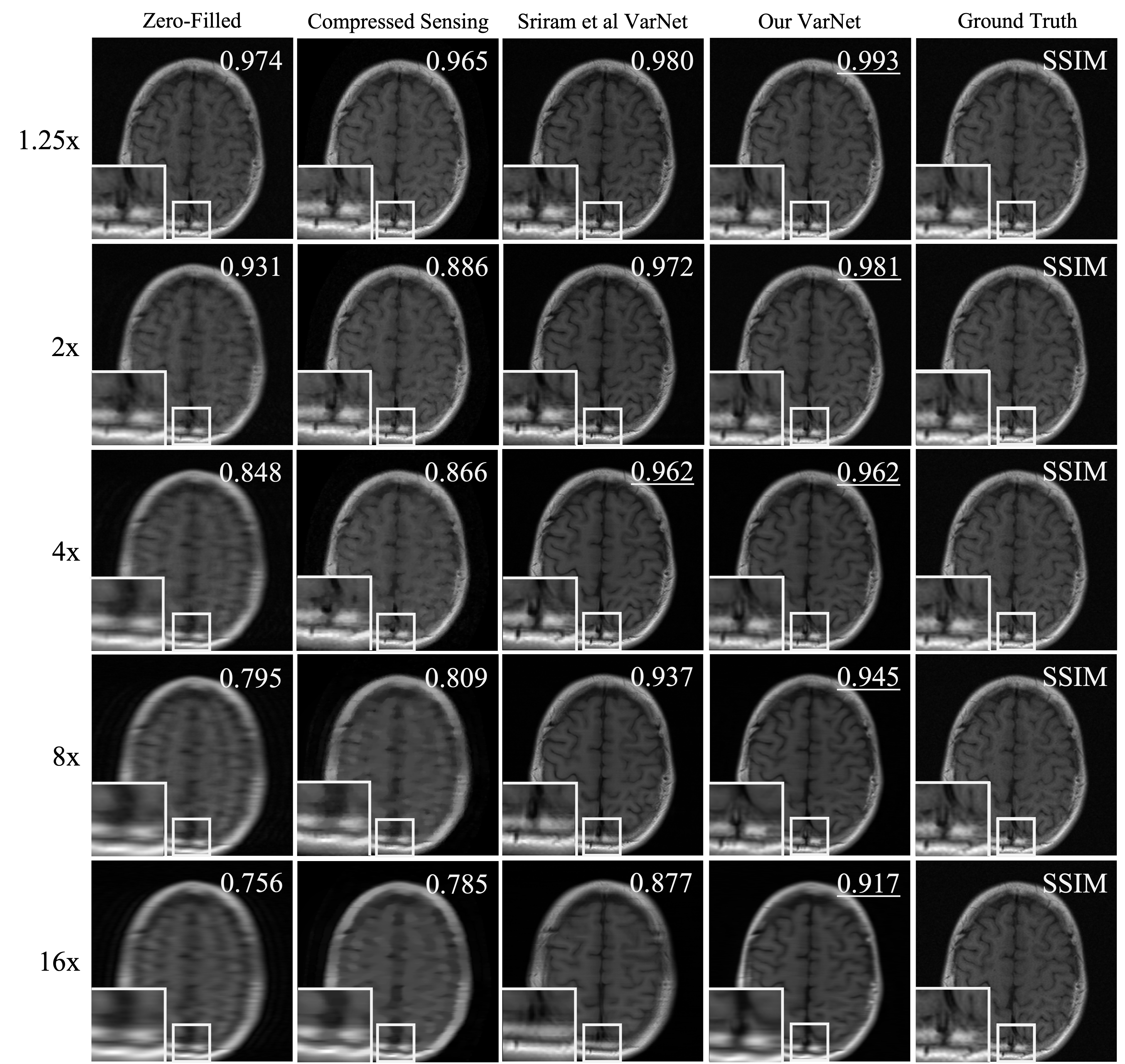}}
\end{figure*}

\begin{figure*}[htpb]
  \floatconts
    {fig:var-knee-histogram}
    {\caption{Histograms of pair-wise SSIM improvements stratified by acceleration factor between an E2E-VarNet model trained on variably accelerated data and one trained on only 4x and 8x accelerated data. The percentages in each plot indicate the proportion of test slices that achieved a higher SSIM metric score using the model trained on variably accelerated data.}}
    {\includegraphics[width=\textwidth]{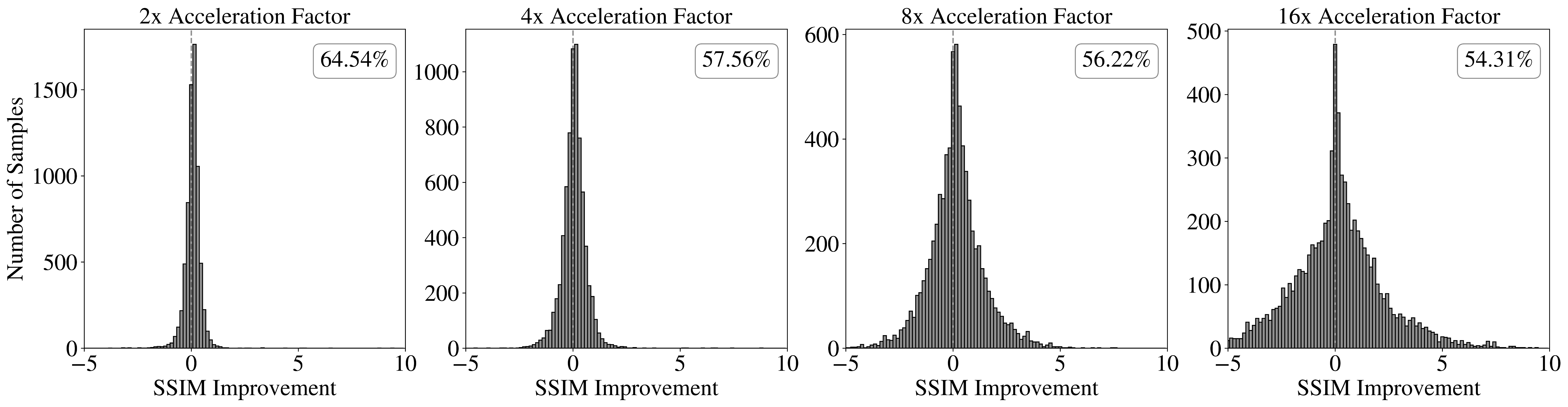}}
\end{figure*}

\begin{figure*}[t!]
  \floatconts
    {fig:tl-histogram}
    {\caption{Histograms of pair-wise SSIM improvements stratified by acceleration factor between an E2E-VarNet model pre-trained on either simulated Shepp-Logan phantoms and one pre-trained on the fastMRI Mini dataset. Both models were fine-tuned on the fastMRI Mini dataset. The percentages in each plot indicate the proportion of test slices that achieved a higher SSIM metric score using the model pre-trained on simulated phantom data.}}
    {\includegraphics[width=0.5\textwidth]{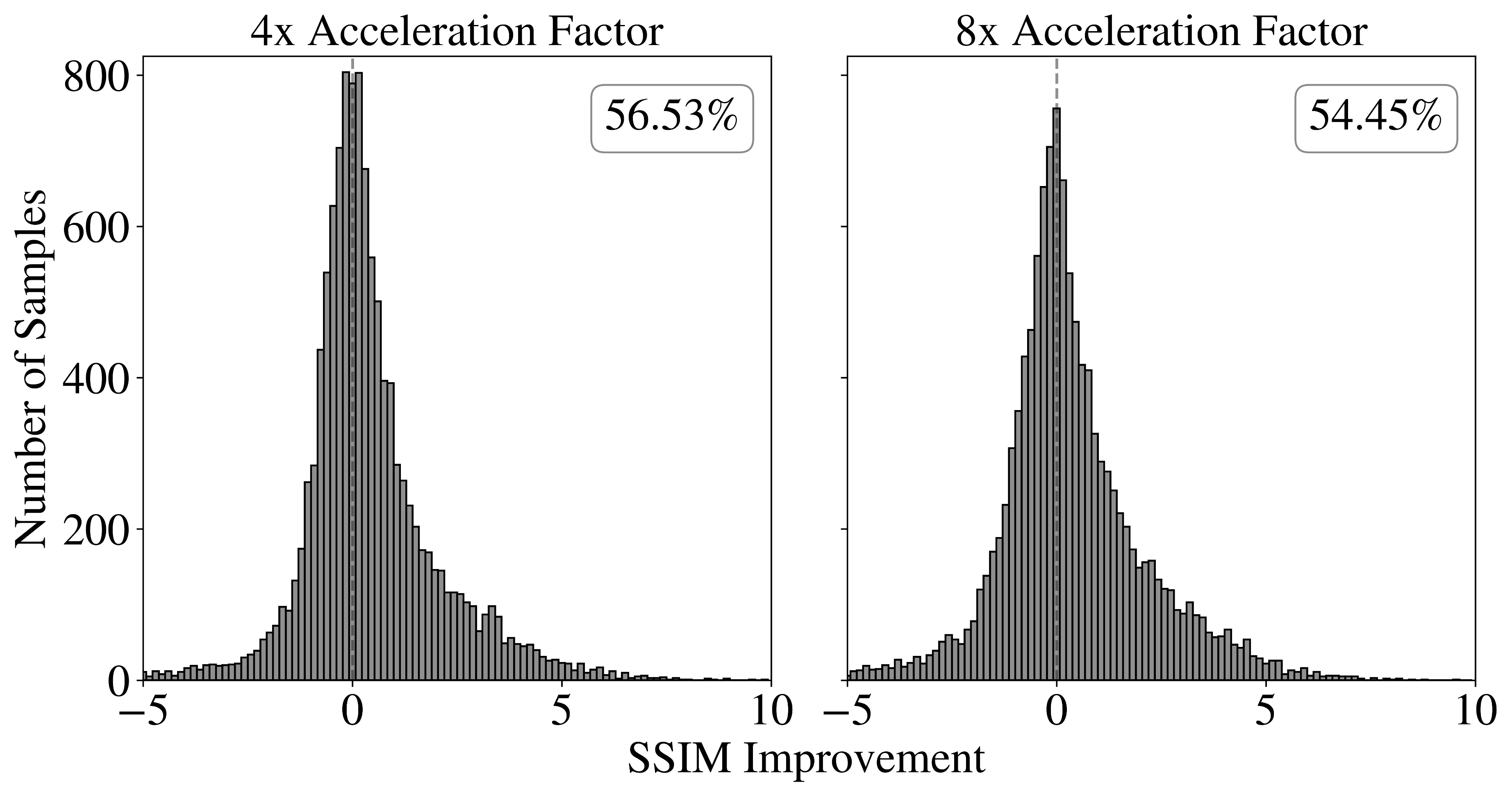}}
\end{figure*}

\subsection{Acceleration Factor Balancing}
\label{supp:reconstructor}
In \tableref{table:reconstructor-brain-supp}, we report reconstructor results exploring the impact of training reconstructor models on variably accelerated undersampled \textit{brain} datasets. We also provide sample image reconstruction panels from our experiments in \sectionref{section:var-acceleration} to illustrate conclusions drawn from our experiments. Images were reconstructed using knee (\figureref{fig:reconstructor-knee}) or brain (\figureref{fig:reconstructor-brain}) models that were both trained on variably accelerated data. Finally, we include a panel of histograms in \figureref{fig:var-knee-histogram} visualizing the distribution of pair-wise SSIM improvement in undersampled multi-coil knee reconstruction using a reconstructor trained on variably accelerated data when compared to one trained on only 4x and 8x accelerated data.

\subsection{Learning Multiple Anatomies}
\label{section:ewc-supp}
\figureref{fig:ewc-by-epoch} demonstrates empirical evidence of catestrophic forgetting in our experiments. As described in \sectionref{section:ewc}, we first trained a reconstructor model to reconstruct undersampled knee data, which was sequentially followed by training on brain data. After each training epoch on the brain dataset, we evaluated the reconstructor's performance on the previously learned knee anatomy. Our results suggest that without EWC regularization, the agent forgets knee anatomy reconstruction within the first epoch of brain training. Furthermore, the consequent decrease in knee SSIM becomes more significant as the acceleration factor increases (\figureref{fig:ewc-images}).

\subsection{Phantom Pre-Training}
\label{section:tl-supp}
\figureref{fig:tl} includes a sample panel of image reconstructions on the final test dataset after model training. Qualitatively, we found that phantom-based pre-training allowed the model to generalize better to previously unseen contexts, such as through reduction of image feature blurring and tissue texture contrast when compared to training on the fastMRI Mini dataset alone (\figureref{fig:tl}). \figureref{fig:tl-histogram} shows a panel of histograms of pair-wise SSIM improvements comparing a reconstructor model pre-trained on simulated phantom data and one pre-trained on the fastMRI Mini dataset.

\end{document}